\documentclass{arxivMS}
\usepackage{booktabs}
\usepackage{caption}
	\shorttitle{Nonlinear compression of sub-nanosecond pulses}                                   
	\shortauthor{Beaufort et al.}
	
	\title{Towards direct nonlinear compression of energetic sub-nanosecond pulses to the ultrafast regime}
	
	\author[1,2,3,*]{Gaspard Beaufort}
	\author[1,4,5]{Nayla Jimenez}
	\author[6]{Gunnar Arisholm}
	\author[1]{Victor Hariton}
	\author[1]{Ayhan Tajalli}
	\author[1]{Ingmar Hartl}
	\author[2]{Anne-Lise Viotti}
	\author[1,4,5,*]{Marcus Seidel}
	
	\address[1]{Deutsches Elektronen-Synchrotron DESY, Notkestr. 85, 22607 Hamburg, Germany}
	\address[2]{Department of Physics, Lund University, P.O. Box 118, SE-22100 Lund, Sweden}
	\address[3]{Department of Physics, Ecole Normale Supérieure de Lyon, 69364, Lyon, France}
	\address[4]{Helmholtz-Institute Jena, Fröbelstieg 3,07743 Jena, Germany}
	\address[5]{GSI Helmholtzzentrum für Schwerionenforschnung GmbH, Planckstrasse 1, 64291 Darmstadt, Germany}
	\address[6]{FFI (Norwegian Defence Research Establishment), P. O. Box 25, NO-2027 Kjeller, Norway}
	
	\correspond{gaspard.beaufort@fysik.lu.se; marcus.seidel@desy.de}
	
	\setabstract{Applications of terawatt-class lasers can enormously benefit from pulse trains with kHz repetition rates. The associated unprecedented combinations of peak and average powers require the development of new concepts for scalable ultrashort pulse generation. We propose using multi-mirror multi-pass cells as a compact and cost-efficient solution for the direct post-compression of sub-nanosecond pulses into the femtosecond regime. We simulate spectral broadening of 100-mJ, 300-ps pulses to a sub-ps Fourier transform-limit in a 1-m diameter multi-pass cell. Furthermore, an 11-mirror cell for about 300 passes was set-up for proof-of-concept. To reach a spectral broadening factor of 15 in air, only 260 MW of peak power were required. The proposed scheme can efficiently transform industrially mature high-power, high-energy lasers into unique ultrafast sources. }
	
\begin{document}
	\maketitle


	
	\section{Introduction}
	\subsection{Extreme nonlinear pulse compression concept}
	Since the invention of chirped pulse amplification (CPA), the  peak powers and intensities of ultrahort laser pulses have continuously been increased, reaching today at large scale facilities up to petawatts and $>10^{22}$\,W/cm$^2$, respectively \cite{mourou_nobel_2019}. The most intense lasers run however only at very low shot rates, which restricts their applicability. Therefore, the concurrent scaling of both peak and average power has been proclaimed as the main objective of the so-called "third generation femtosecond technology" \cite{fattahi_third-generation_2014}. For this purpose, the conventional laser materials for ultraintense light sources, namely Ti:sapphire and large aperture Nd:doped glasses, are very difficult to use because of their unfavorable thermal properties. Alternatively, optical parametric chirped pulse amplifiers (OPCPAs) are used to minimize heat generation in the amplifier crystals at the expense of zero storage time and moderate quantum efficiencies \cite{fattahi_third-generation_2014}.\par
	High average power OPCPAs typically use Yb-based amplifiers or their harmonics as pump sources. Yb-lasers offer excellent average power scalability through fiber, innoslab or thin-disk geometries \cite{zuo_highpower_2022}. But it is very difficult to extract Joule-level pulse energies since Yb:YAG is a quasi three-level system. For thin-disk amplifiers, which enabled 1\,kHz repetition rate ultrashort pulses with up to 720 mJ pulse energy \cite{herkommer_ultrafast_2020}, the low single pass gain poses an additional challenge.\par
	\begin{figure}[t]
		\centering
		\includegraphics[width=\linewidth]{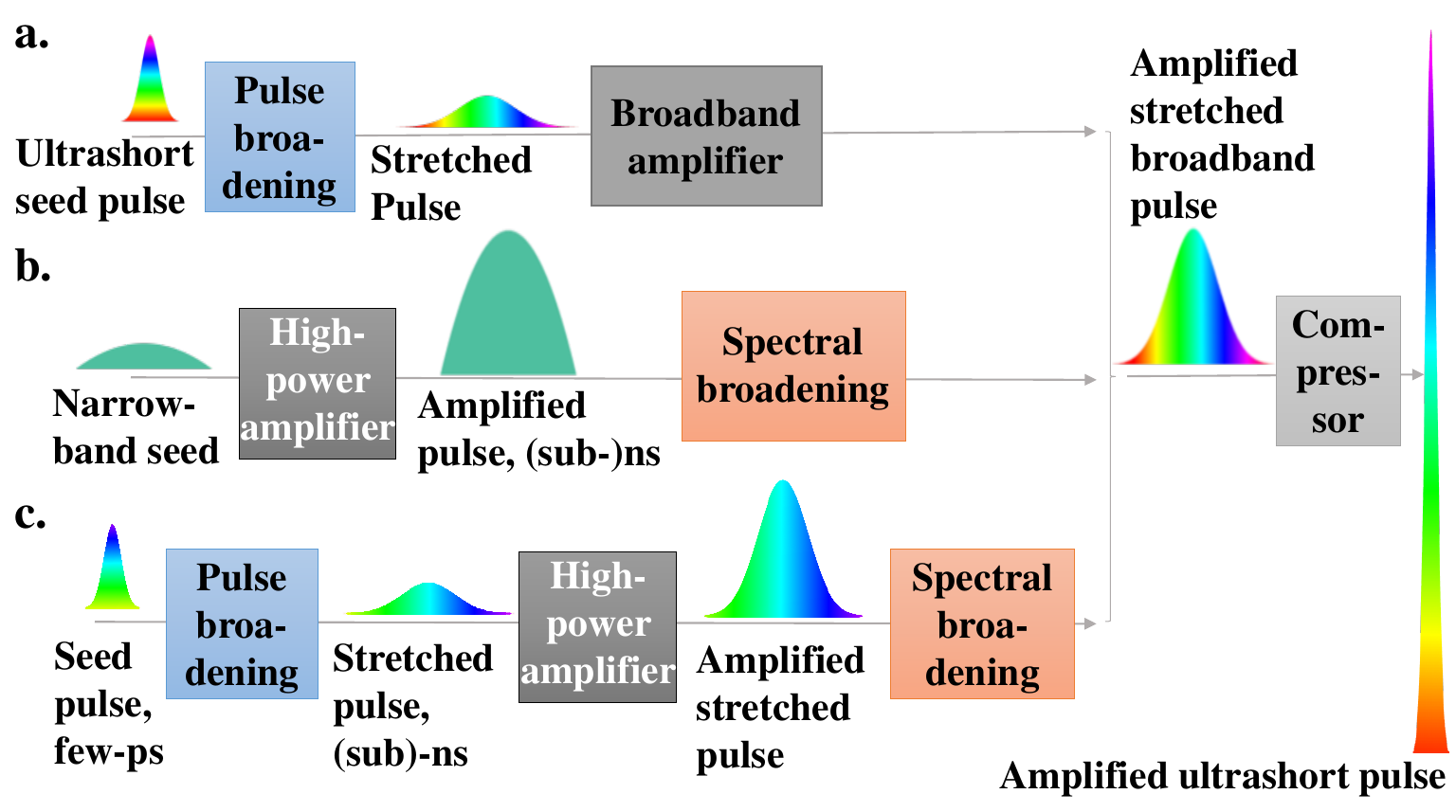}
		\caption{\textbf{a.} Conventional CPA scheme: A broadband, ultrashort seed pulse is stretched by spectral domain chirping, amplified and recompressed. \textbf{b.} Proposed CAA scheme: A narrowband, long seed pulse is first amplified, then spectrally broadened by time domain chirping and finally recompressed. The scheme can be applied e.g. to Nd:YAG lasers. \textbf{c.} Hybrid scheme. Pulse broadening before amplification leads to moderate intensities in the amplifiers. Spectral broadening before compression enables compact setups. The scheme can be applied e.g. to cryogenically cooled Yb:YAG lasers. }
		\label{fig1:CAA}
	\end{figure}
	Here, we propose a novel ultrashort pulse generation scheme which relies on gain media that provide both the capability to generate Joule-level pulse energies and to handle kilowatts of average powers, e.g. Nd:YAG \cite{fan_high_2017}, cryogenically cooled Yb:YAG \cite{wang_1J_1kHz_cryo_ybyAG_2020,vido_10J_100Hz_dipole_2024} or potentially CO$_2$ \cite{tochitsky_prospects_2016}. These materials offer only narrow gain bandwidths in comparison to state-of-the-art ultrafast lasers, and thus cannot provide femtosecond pulses without further spectral broadening. The paper will show that multi-mirror multi-pass cells (MM-MPCs) present a viable platform for attaining multiple hundredfold pulse bandwidth extensions in order to turn conventional long pulse lasers into ultrafast sources. The concept is in many regards similar to CPA. The difference is that CPA requires pulse broadening, i.e. spectral chirp, before amplification, while the proposed scheme requires spectral broadening, i.e. temporal chirp, after amplification. We therefore call the approach "chirp after amplification" (CAA). The analogies are shown in Figure~\ref{fig1:CAA}.\par
	
	\subsection{Implementation approach: spectral broadening in multi-pass cells}
	For short optical pulses, spectral broadening is efficiently achieved by virtue of the nonlinear optical Kerr effect resulting in self-phase modulation (SPM). Due to losses and pulse broadening by dispersion, SPM saturates in the normal dispersion regime. One can derive an optimal spectral broadening factor $\beta_\text{opt}$ and corresponding propagation length $L_\text{opt}$ resulting in a good compromise between pulse duration and pulse quality \cite{tomlinson_compression_1984,diels_pulse_2006}. For a virtually lossless medium, and under the assumptions of linear dispersion and an instantaneous nonlinear response, one finds \cite{tomlinson_compression_1984,diels_pulse_2006}:
	\begin{subequations}
		\begin{equation}
			\beta_\text{opt} \propto \sqrt{L_D/L_{NL}} \propto \Delta t_0.
			\label{eq:b_opt}
		\end{equation}
		\begin{equation}
			L_\text{opt} \propto \sqrt{{L_D}{L_{NL}}} \propto \Delta t_0,
			\label{eq:L_opt}
		\end{equation}
	\end{subequations}
	where $L_D$ and $L_{NL}$ are the dispersive and nonlinear lengths, respectively, and $\Delta t_0$ is the input pulse duration. The linear dependence of $\beta_\text{opt}$ on $\Delta t_0$ originates from the reduced pulse sensitivity of longer pulses to dispersion. On the other hand, the likewise reduced temporal gradient of the pulse slows down the bandwidth gain during nonlinear propagation. The physical propagation length $L_\text{opt}$ required to reach $\beta_\text{opt}$ is thus also proportional to $\Delta t_0$. 
	This implies that 300\,ps pulses would have to propagate a hundred times longer than 3\,ps pulses to reach the same Fourier transform (FT) limit after spectral broadening, assuming the same $L_{NL}$  for both cases. Therefore, scaling spectral broadening by orders of magnitude requires power-scalable, low-loss waveguides.\par 
	With the recent emergence of the multi-pass cell (MPC) spectral broadening platform \cite{schulte_nonlinear_2016} and the rapid advance of low-loss anti-resonant hollow-core fibers \cite{fokoua_loss_2023}, implementing the CAA concept has come in sight. We only briefly discuss the potential of hollow-core fibers in appendix \ref{app1:hc-fiber} since the transmitted pulse energies reported to date are only in the few-mJ range. By contrast, operation at $>100\,$mJ pulse energies has been experimentally verified with MPCs \cite{kaumanns_spectral_2021,pfaff_nonlinear_2023}. They enable nonlinear propagation over long distances while maintaining high transmission efficiency. Highly-reflective mirrors reach reflectances of $R > 99.999\,\%$, and thus the propagation can be virtually lossless. MPCs naturally operate in the normal dispersion regime and gas-filled cells exhibit only small group velocity dispersion. Moreover, by using atomic gases, Raman and Brillouin scattering are eliminated. Finally, MPCs have already proven to operate at kilowatt average as well as multi-ten-gigawatt peak powers and reached compression factors larger than 30 with 95\,\% power efficiency \cite{viotti_multi-pass_2022}. Consequently, MPCs present the ideal platform for CAA.  
	Here, two decisive steps towards implementing the concept are reported. First, the SISYFOS simulation package \cite{arisholm_simulation_2021} is used to investigate effects that go beyond basic spectral broadening theory. Second, multi-mirror multi-pass cells are used for the first time for spectral broadening experiments. MM-MPCs enable much larger nonlinear propagation lengths than usual Herriott-type MPCs and will be essential for the nonlinear compression of energetic narrow-band pulses.\par
	
	
	
	\section{Simulation: Spectral broadening of 100\,mJ, 300\,ps pulses in multi-pass cells}
	\subsection{Simulation settings}
	
	We used the SISYFOS package, which has excellently modeled multiple MPC experiments \cite{seidel_factor_2022,viotti_few-cycle_2023}, and numerically studied spectral broadening of 300 ps (FWHM), 100 mJ pulses over 1300 MPC passes. The pulse parameters were chosen because they are readily accessible by cryogenically cooled Yb:YAG and commercially available Nd:YAG lasers. \par 
	The multi-pass cell was modeled under the following approximations: The beam was always on the optical axis of the system. The input beam was a fundamental Gaussian ($M^2 = 1$). Reflectivity and dispersion of the MPC mirrors, essentially SiO$_2$/Ta$_2$O$_5$ quarter-wave-stacks, were taken from theoretical data provided by the high-reflectivity mirror supplier. The MPC mirrors had 500\,mm radii of curvature ($\mathcal{R}$) and were separated by 975\,mm. Their reflectivity of 99.99\,\% resulted in 87\,\% transmission after 1300 bounces. The fluence on the mirrors was below 6\,J/cm$^2$. The Kerr medium was krypton. Its nonlinear refractive index was derived from ref.~\cite{travers_ultrafast_2011}, the dispersion from ref.~\cite{borzsonyi_measurement_2010}. Up to $2^{14}$ points along the time axis were used. This was required due to the large broadening factors for the highest gas pressures studied. The simulations automatically included self-steepening by solving the nonlinear wave equation in the frequency domain. However, due to the sub-ns pulse durations, self-steeping was not observed.\par
	
	\begin{figure*}[t]
		\centering
		\includegraphics[width=\linewidth]{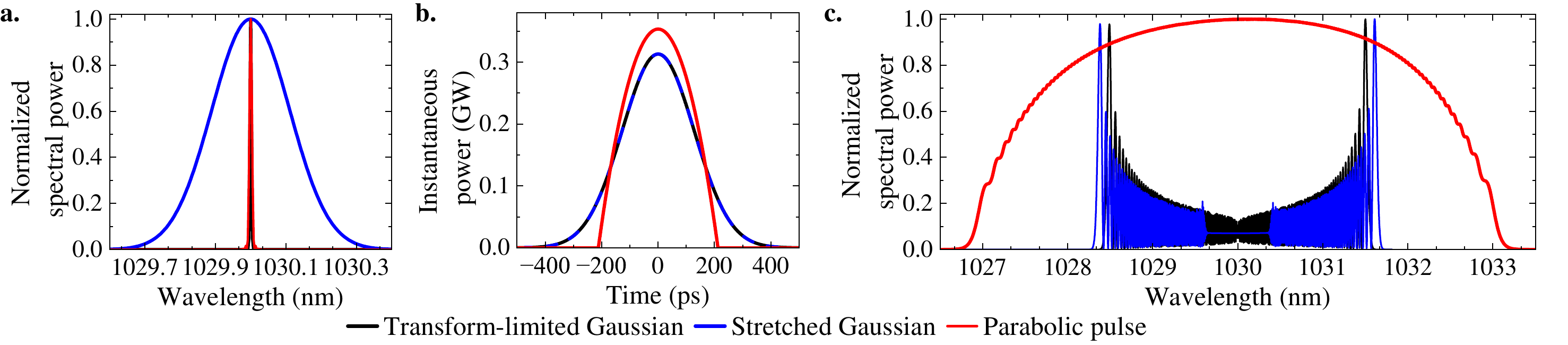}
		\caption{Simulated spectral broadening of 300\,ps, 100\,mJ input pulses in a 1300-pass MPC for Fourier-transform (FT) limited pulses with Gaussian (black lines) and parabolic intensity envelope (red lines) as well as  stretched Gaussian pulses with 6\,ps FT limit (blue lines). \textbf{a.} Input spectra of the simulations. \textbf{b.} Input pulses of the simulations. The pulse shape is virtually preserved during propagation. \textbf{c.} Simulated output spectra for a Kr pressure of $p=4.5\,$bar.}
		\label{fig2:CAA_sim}
	\end{figure*}
	
	\subsection{Input pulse dependent spectral broadening}
	Following the pictorial scheme of Figure~\ref{fig1:CAA}, we investigated different input pulse shapes and spectra.  For this purpose, simulations with cylindrical symmetry and a grid with 256 points along the radial coordinate were conducted. We first considered a FT limited Gaussian pulse with the complex electric field: 
	\begin{equation}
		E_{G}(t) = E_0\exp\left[-2\ln2(t/\Delta t_p)^2+i\omega_ct\right],
		\label{eq:Gauss}
	\end{equation}
	and a positively chirped Gaussian pulse with 6\,ps FT limit,
	\begin{align}
		&E_{sG}(t) = \sqrt{\frac{\pi E_0}{2\ln 2}}\Delta t_{p0}\,\times \nonumber \\
		&\mathcal{FT}\left\{\exp\left[\frac{\Delta t_{p0}^2(\omega-\omega_c)^2\left(1+i\left[\frac{\Delta t^2_p}{\Delta t^2_{p0}}-1\right]^{1/2}\right)}{8 \ln 2}\right]\right\},
		\label{eq:strGauss}
	\end{align}
	where the field amplitude $E_0$ was set to obtain 100\,mJ pulse energy, $\Delta t_p=300\,$ps, $\omega_c=2\pi\cdot 299.8\,\frac{\mu\text{m}}{\text{ps}} / 1.03\, \mu\text{m}$ and $\Delta t_{p0} = 6\,$ps. $\mathcal{FT}$ denotes the Fourier transformation from the frequency to the time domain. Input pulses and spectra for the simulations are shown in Figure~\ref{fig2:CAA_sim}a and b.\par 
	The cell was filled with 4.5\,bar of krypton, and thus the peak power was about 17\,\% of the critical power. The broadened spectra plotted as black and blue lines in Figure~\ref{fig2:CAA_sim}c correspond to sub-picosecond FT limits of 896\,fs and 837\,fs, respectively, but they are strongly modulated. Only the spectral range of the input spectrum remains smooth in the output spectrum. Consequently, a main challenge for implementing the scheme is the minimization of the non-quadratic phase, which Gaussian pulses acquire from SPM. It limits the peak power enhancement and pulse contrast at high compression factors. This can be circumvented by shaping the input pulses to a parabolic intensity profile, in our simulations described by:
	\begin{equation}
		E_{P}(t) = E_0\,Re\left\{\sqrt{1-2(t/\Delta t_p)^2}\right\}e^{i\omega_ct},
		\label{eq:parabol}
	\end{equation}
	and depicted by the red lines in Figure~\ref{fig2:CAA_sim}. Owing to the quadratic intensity envelope, SPM induces quadratic phase in analogy to the dominant dispersion term of pulse stretchers in CPA schemes. The red line in Figure~\ref{fig2:CAA_sim}c shows the corresponding smooth output spectrum with 546\,fs FT limit. Several approaches to generate parabolic pulses have been reported, e.g. in refs. \cite{finot_optical_2009,weiner_ultrafast_2011,liu_generation_2019}. Investigating their applicability to our high-power pulse generation concept will be subject of follow-up research.\par
	It is important to emphasize that effects different from SPM, e.g. optical wave-breaking, self-steepening or modulational instabilities, are absent even for $>1000$ spectral broadening factors which are reached at higher krypton pressures. This is enabled by the combination of long input pulses and MPCs as spectral broadening platform. 
	
	\begin{figure*}[t!]
		\centering
		\includegraphics[width=\textwidth]{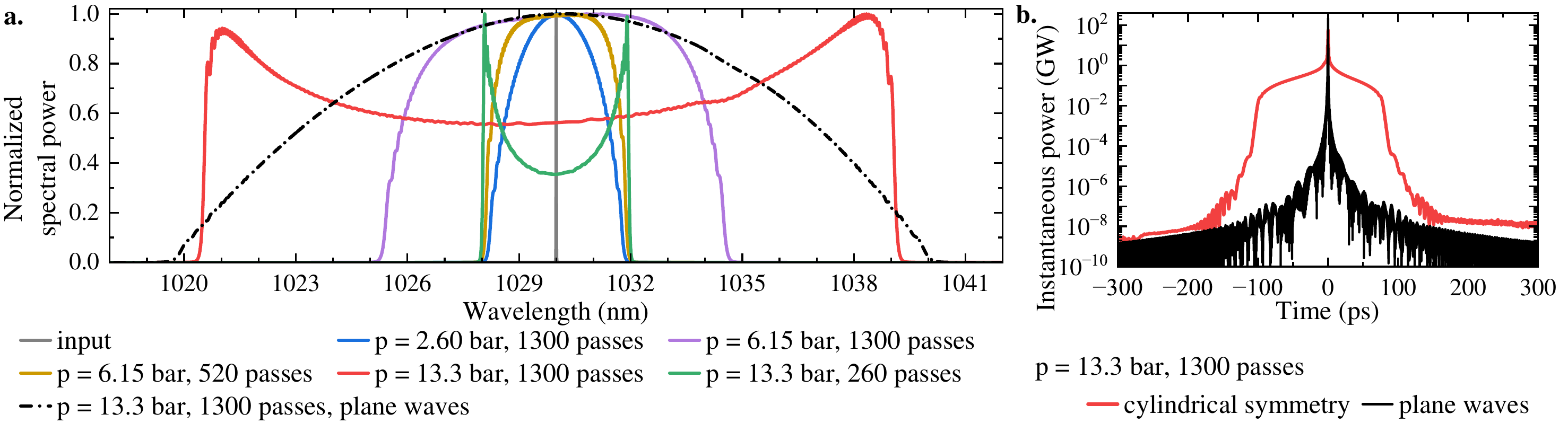}\\
		\caption{The simulations presented in this figure use the parabolic input pulse described by \eqref{eq:parabol}. \textbf{a.} The solid lines show a comparison between output spectra after 1300 passes for $p = p_{cr}/2$ (red), $p_{cr}/4$ (violet) and $p_{cr}/10$ (blue). Spatial effects are included in the simulations by assuming cylindrical symmetry. The black dashed line results from a simulation with only a single spatial grid point (i.e. plane waves). The krypton pressure was also set to $p = p_{cr}/2$. The comparison to the red solid line shows a clear impact of spatio-temporal couplings on the spectral shape. These couplings scale with the peak-to-critical power ratio as the comparison between the green, ocher and blue lines shows, which exhibit the same bandwidth but different spectral shapes. The curves were obtained by adjusting the number of passes and krypton pressures, such that comparable B-integrals were acquired in all cases. The pressure is inversely proportional to the critical power of the nonlinear gas. \textbf{b.} Best pulse compression to about 400\,fs duration is achieved with a group delay dispersion of  GDD = -7.4\, ps$^2$ and a third-order dispersion of TOD=-0.021 ps$^3$ for the simulation assuming cylindrical symmetry. The compression result is not very sensitive to TOD. Higher-order dispersion terms are not considered. A group delay dispersion of GDD = -11.9\, ps$^2$ and a third-order dispersion of TOD=-0.036 ps$^3$ resulted in nearly FT limited pulses with 176\,fs duration and excellent contrast for the plane wave simulation. The result is more sensitive to TOD optimization.}
		\label{CAA_sim_press}
	\end{figure*}
	
	\subsection{Role of sub-critical self-focusing}
	\label{subsec:self-focusing}
	As both $L_D$ and $L_{NL}$ scale inversely with gas pressure, high pressures are favorable to keep the propagation length as short as possible (Eq.~\eqref{eq:L_opt}). In ref. \cite{viotti_multi-pass_2022}, the pressure limit was assumed to be set by the equality of peak power $P_p$ and critical self-focusing power $P_{cr}$. In practice, MPC experiments with pressures up to $P_p = 0.4\,P_{cr}$ have been reported \cite{rajhans_post-compression_2023}. For krypton as nonlinear medium and the laser parameters we investigated, a pressure of  $p_{cr} = 26.6\,$bar results in $P_p = P_{cr}$. However, sub-critical self-focusing changed the shape of the SPM broadened spectra significantly also for $p = p_{cr}/2 = 13.3\,$bar and $p = p_{cr}/4 = 6.15\,$bar as the red and violet lines in Fig.\ref{CAA_sim_press}a show. For comparison, the dashed black line results from a plane wave simulation with $p = p_{cr}/2$, i.e. without spatial degrees of freedom. Its bell shape is in clear contrast to the red solid line obtained with identical nonlinear and dispersive lengths, but by additionally including the radius $r$ as a transversal coordinate in the simulation. The broadened spectrum for $p = p_{cr}/10 = 2.66\,$bar (Figure~\ref{CAA_sim_press}a blue line) does not exhibit significant features of self-focusing, but resembles the shape of a plane wave simulation. With reduced numbers of passes, comparable broadening factors were simulated with $p = p_{cr}/4$ and $p = p_{cr}/2$. The spectra are represented by the green and ocher lines in Figure~\ref{CAA_sim_press}a. These also indicate nonlinear space-time couplings. Along with the change of spectral shape, a change of spectral phase was observed. Figure~\ref{CAA_sim_press}b shows that the compressibility of the pulses may suffer from space-time-couplings. While the self-phase modulated pulses could be compressed close to their FT limit of 153\,fs in the case of plane wave simulations (black line), only a pulse duration of about 400\,fs and a bad pulse contrast could be computed if a transversal degree of freedom was included (red line). The impact of spatial-temporal couplings gets even more pronounced if astigmatism is introduced as an additional degree of freedom, which is discussed explicitly in appendix~\ref{app2:astig}.
	To our knowledge, phase distortions due to self-focusing have been investigated for bulk-based MPCs \cite{seidel_ultrafast_2022}, but comparable observations for gas-filled MPCs are lacking. On the contrary, in studies with $P_p>0.4 P_{cr}$, 24-fold compression to a near FT limited pulse duration of 50\,fs with a good pulse contrast was reported \cite{rajhans_post-compression_2023}. Therefore, the numerical predictions require additional experimental investigations.

	\subsection{Pulse compression}
	\label{subsec:sim_compress}
	\begin{figure*}[t!]
		\centering
		\includegraphics[width=\textwidth]{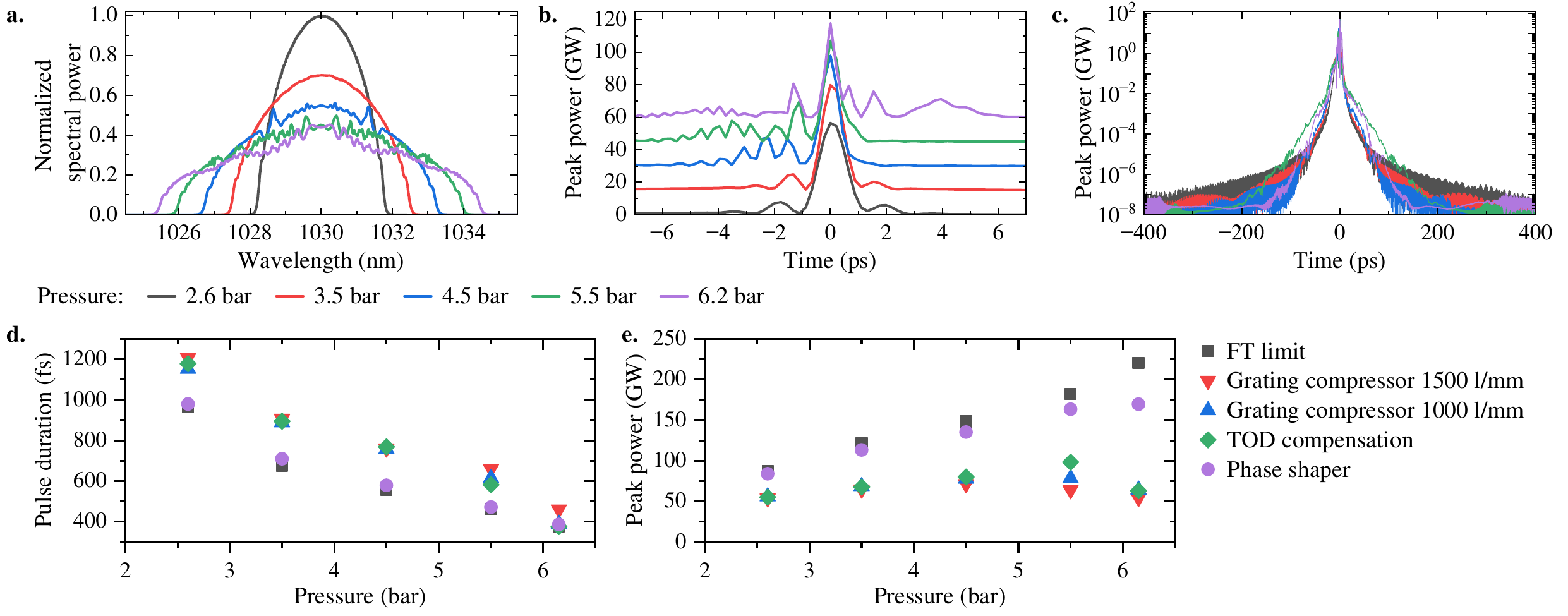}
		\caption{Simulated spectral broadening of pulses with parabolic envelop in dependence of the krypton pressure in an 11-mirror MPC. 1300 passes were simulated. \textbf{a.} Simulated spectra after 1300 passes. \textbf{b.} Post-compressed pulses after dispersion compensation with a 1500 lines/mm grating pair compressor. The pulses are offset by 15 GW for clarity. \textbf{c.} Compressed pulses plotted on a logarithmic scale. \textbf{d.} Overview of simulated output pulse durations for the different compressor types. \textbf{e.} Overview of simulated output peak powers for the different compressor types. }
		\label{fig:sim_compress}
	\end{figure*}
	
	\begin{table}[b!]
		\centering
		\caption{Simulated pulse compressors}
		\begin{tabular}{l c c}
			\toprule
			& GDD & TOD \\
			\midrule
			1500 lines/mm gratings* & $-34\,\text{ps}^2/m$ & $0.221\,\text{ps}^3/m$  \\
			1000 lines/mm gratings* & $-6.14\,\text{ps}^2/m$ & $0.0174\,\text{ps}^3/m$  \\
			TOD-free compressor & optimized & $-1.4\cdot 10^{-4}\,\text{ps}^3$  \\
			Phase shaper** & best fit & best fit   \\
			\bottomrule
		\end{tabular}
		\label{tab:compress}
		\caption*{* unit $m$ refers to grating pair separation; ** higher-order dispersion up to $6^\text{th}$ order is also fitted}
	\end{table}
	
	{In regard of the observed space-time couplings, we used pulse compression simulations to derive an optimized critical-to-peak-power ratio. Four compressor types were studied: Two grating pair compressors using 1500 lines/mm and 1000 lines/mm gratings in Littrow configuration. Moreover, a compressor compensating the small TOD of the MPC mirrors ($106\,\text{fs}^3$ per bounce) and the linear chirp of the pulses was considered and eventually, a phase shaper with control of up to the 6$^\text{th}$-order dispersion term was applied. An overview of the compressors is presented in Table~\ref{tab:compress}. Spectral broadening with krypton pressures of $p = 2.66$\,bar ($p = p_{cr}/10$), $p = 3.5$\,bar, $p = 4.5$\,bar, $p = 5.5$\,bar and $p = 6.15$\,bar ($p = p_{cr}/4$) was simulated including the astigmatism induced by the MPC mirrors (cf. appendix~\ref{app2:astig}). The spectra after passing through the MPC are shown in Figure~\ref{fig:sim_compress}a. To optimize the grating distances and the phase shaper, respectively, only the residual spectral phase of the x=0, y=0 simulation grid point was considered. For the grating compressors, the grating distance was optimized with respect to peak power. For the phase shaper, the spectral phase was fitted by a polynomial. Appendix~\ref{app3:compress} shows the redidual spectral phases for compression the $p = 2.66$\,bar and $p = 6.15$\,bar pressure cases.  The derived pulse at the central beam position agreed very well with the beam averaged pulse, which indicates the excellent spatial-spectral homogeneity known from MPC spectral broadening.\par 
		An overview of the simulation results is presented in Figure~\ref{fig:sim_compress}. For all pressures, the pulses can be compressed relatively close to their FT-limit (Figure~\ref{fig:sim_compress}d). However, the power in the pulse pedestals increases considerably with pressure unless the phase shaper is used for compression. As a consequence, the highest peak power enhancements are obtained at pressures of 4.5\,bar or 5.5\,bar if higher-order dispersion terms are not compensated (Figure~\ref{fig:sim_compress}b,e). The compression is most sensitive to TOD for $p = 5.5$\,bar. In this case, the peak power can be increased from 64 \,GW to 79\,GW bei reducing the line density of the gratings from 1500/mm to 1000/mm and to 98\,GW if mirrors and gratings were TOD-free. Figure~\ref{fig:sim_compress}c shows that even for the 1500 line/mm grating compressor pair, the instantaneous power drops quickly on both sides of the main pulses to the $10^{-9}$ level relative to the peak. A broad pedestal like in Figure~\ref{CAA_sim_press} is not visible. We also note that the amplified stimulated emission background is strongly suppressed by the nonlinear peak power enhancement and the eigenmode properties of MPCs.\par
		In summary, our simulations predict that the best operation point for the MM-MPC experiments is at peak power levels around 20\,\% of the critical power of the nonlinear gas. At this point, 500-fold compression from 300\,ps to 600\,fs is predicted if TOD is at least partially compensated. Higher-order phase control appears clearly favorable for further peak power enhancement and the reduction of pulse pedestals in the few-ps time range. These pedestals could also be cleaned by nonlinear ellipse rotation \cite{seidel_ultrafast_2022,pfaff_nonlinear_2022} in a second nonlinear compression stage that would enable TW peak powers and pulse durations of 30\,fs - 50\,fs \cite{kaumanns_spectral_2021,pfaff_nonlinear_2023}.  \par

	\section{Multi-mirror multi-pass cells}
	
	\subsection{Design of the multi-mirror multi-pass cell}
	
	The setup compactness, with a mirror separation of less than 1\,m, is enabled by the huge single-stage pulse compression factors. Ultrafast sources with the same pulse energy typically require very long cells \cite{viotti_multi-pass_2022} which are difficult to engineer. However, the simulated MPCs used 1300 reflections off the mirrors, which presents a major experimental challenge. At present, state-of-the art Herriott-type MPCs are operated with 20 – 30 roundtrips for spectral broadening, because they only use a small fraction of the mirror surfaces. Using slightly astigmatic mirrors yields Lissajous spot patterns on the mirrors \cite{herriott_folded_1965}, and thus a significantly larger beam coverage of their surface area. Consequently, the number of passes through an MPC can be strongly increased. We investigated this approach experimentally. One way to introduce astigmatism is to arrange at least three mirrors on a circle, allowing the beam to propagate in the MPC in a star-like path with a non-zero angle of incidence $\varphi\neq0$ on the mirrors (Figure~\ref{fig3:MM-MPC_experiment}a). The focal lengths of the curved mirrors then differ slightly between planes, being $f_{sag}\approx \frac{\mathcal{R}\cos(\varphi)}{2}$ in the sagittal plane and $f_{tan}\approx \frac{\mathcal{R}}{2\cos(\varphi)}$ in the tangential plane \cite{jenkins_fundamentals_1976}. 
	Such setups have recently been studied for absorption spectroscopy applications. To our knowledge, only experiments with up to five mirrors were demonstrated \cite{hudzikowski_compact_2021,kong_optical_2022}. However, to reduce aberrations introduced by off-axis reflections on the spherical mirrors and to increase further the number of reflections, we investigated the design of an eleven-mirror MPC. For predicting the beam paths, Kong et al. \cite{kong_optical_2022} solved the ray propagation in three dimensions using the vector reflection principle and then applied a particle swarm optimization algorithm. We use here an analytical formula based on the ABCD matrix formalism that strongly reduces the computational cost and yields the same propagation results (Appendix~\ref{app5:ABCD}). 
	
	\begin{figure*}[t]
		\centering
		\includegraphics[width=\textwidth]{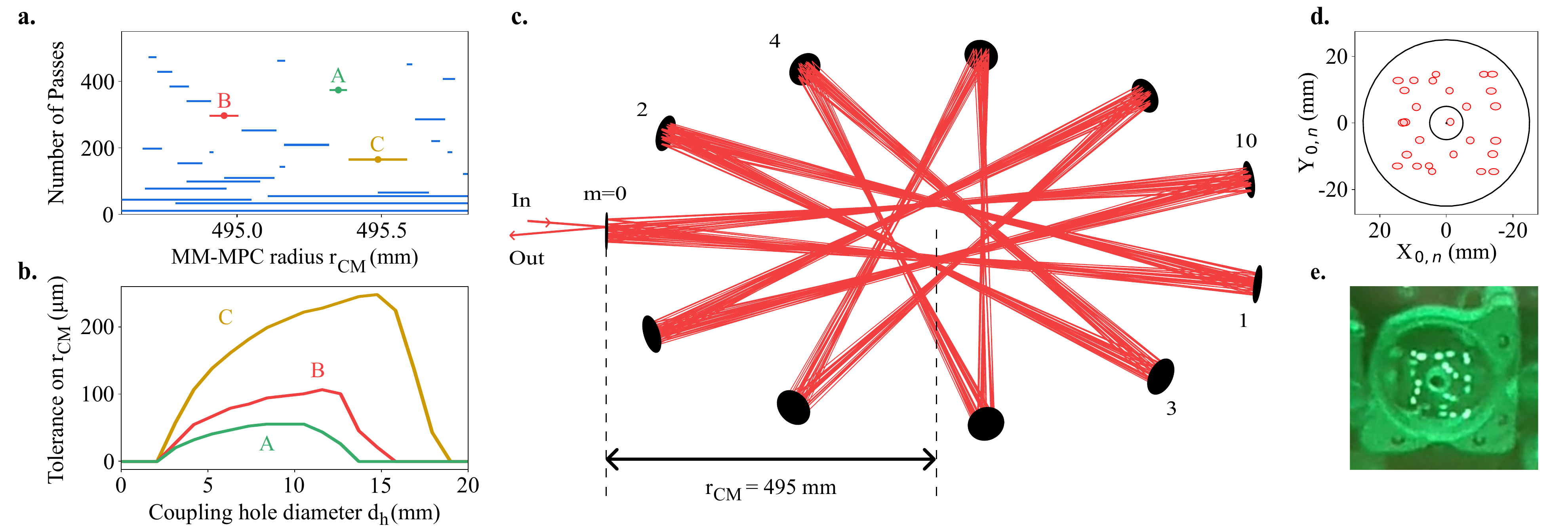}
		\caption{\textbf{a.} Calculated number of round-trips of non-clipping patterns as a function of the MM-MPC radius $r_{CM}$, for a coupling hole diameter $d_h$=10 mm. \textbf{b.} Tolerance on $r_{CM}$ as a function of the hole diameter $d_h$ for three selected patterns A, B and C with 374, 297 and 165 passes, respectively. \textbf{c.} Simulated beam trajectory inside the 11-mirror MPC, for pattern B (297 passes). \textbf{d.} Corresponding simulated reflection pattern on input mirror m=0, showing the central coupling hole. \textbf{e.} Corresponding experimental reflection pattern.}
		\label{fig3:MM-MPC_experiment}
	\end{figure*}
	
	We consider an MM-MPC with $M$ mirrors arranged in a circle. The laser beam is coupled in and out via a hole at the center of mirror $m=0$. In the paraxial approximation, the $X-Y$ coordinates of the $n^{th}$ reflection relative to the center of the mirror $m$ are then:
	\begin{subequations}
		\begin{equation}
			X_{m,n} = X_{\text{max}} \sin[(n M+m) \theta_x + n M \pi],
			\label{eq:Xkm}
		\end{equation}
		\begin{equation}
			Y_{m,n} = Y_{\text{max}} \sin[(n M+m) \theta_y],
		\end{equation}
		\label{eq:XYkm}
	\end{subequations}
	
	where $m=0\ldots M-1$ is the mirror index. $X_{\text{max}}$ and $Y_{\text{max}}$ are constants that only depend on the input angle, while $\theta_x$ and $\theta_y$ are angular advances that are solutions of $\cos{\theta_x} = 1 - \frac{D_{M} \cos{\varphi}}{\mathcal{R}}$ and $\cos{\theta_y} = 1 - \frac{D_{M}}{\mathcal{R}\cos{\varphi}}$, where $D_M$ is the distance between the centers of two consecutive mirrors. The formalism is identical to an MPC with two astigmatic mirrors \cite{herriott_folded_1965} apart from the $nM\pi$-term in \eqref{eq:Xkm}. Only three parameters define the beam pattern: the distance $r_{CM}$ from the center of the MM-MPC to each mirror, and the horizontal and vertical input angles. Beyond that, the diameter of the mirrors $d_m$ and of the coupling hole $d_h$ determine feasible spot patterns since the beam must neither clip during the coupling in and out of the MM-MPC, nor clip on the hole while being reflected from the mirror $m=0$. \par
	
	In order to determine $r_{CM}$ and the input angles that provide fully coupled out patterns, we evaluated all possible trajectories using Eqs.~(\ref{eq:XYkm}a,b). Assuming the mode-matching condition for $\varphi\rightarrow0$, we computed the beam radius on the surface of the MM-MPC mirrors. Then, we selected only patterns in which all the beam spots remained at least one beam radius away from the in-/out-coupling hole, for a given mirror diameter $d_m$ and hole diameter $d_h$ (Figure~\ref{fig3:MM-MPC_experiment}a). For each valid pattern, we define a tolerance on $r_{CM}$ as the range from the minimal to the maximal value of $r_{CM}$ where the pattern can be achieved without clipping. This tolerance can be optimized for a given mirror diameter $d_m$ by appropriately choosing the hole diameter $d_h$, as illustrated in Figure~\ref{fig3:MM-MPC_experiment}b. This optimization helps in designing the coupling hole in order to minimize the sensitivity of the pattern on $r_{CM}$.	
	
	\subsection{Experiment: 11-mirror multi-pass cell}
	\label{sec:11MMMPC}
	In a proof-of-concept experiment, we used eleven spherical mirrors with a $\mathcal{R}=500$\,mm and $d_m=50.8$\,mm (Figure~\ref{fig3:MM-MPC_experiment}c). Numerical analysis did not yield any pattern supporting 1300 passes with this mirror size. As no larger mirrors were available at the time of the experiment, we instead targeted a pattern with 27 reflections per mirror, resulting in a total of 297 passes (Figure~\ref{fig3:MM-MPC_experiment}d). While the beam paths through the MM-MPCs can be readily computed, the experimental implementation requires the control over $5 \times M+2$ degrees of freedom, including the input beam angles and the position and the angle of each mirror. Fortunately, the degrees of freedom can be mostly decoupled. First, the mirrors were placed using a precision rotation stage at the MM-MPC center connected to an adjustable rod. In order to set the angle of the mirrors, the input beam was centered on each mirror, i.e. $X_{\text{max}} = Y_{\text{max}}=0$. This was achieved with high precision by generating first vertical ($X_{\text{max}}=0$) and second horizontal ($Y_{\text{max}}=0$) beam patterns by changing the input angles. By comparing beam patterns for $X_{\text{max}} \neq 0$ and $Y_{\text{max}}\neq 0$ with numerical results, an estimation of $r_{CM}$ can be extracted. The length of the rod was then fine-tuned to the target position thanks to a translation stage. The procedure was then repeated until the targeted pattern with 297 passes was obtained. Figs.~\ref{fig3:MM-MPC_experiment}d and e show good agreement between the simulated and the experimentally obtained beam patterns.  \par

	\subsection{Patterns from 3-inch diameter mirrors}
	
	\begin{figure}[t]
		\centering
		\includegraphics[width=\linewidth]{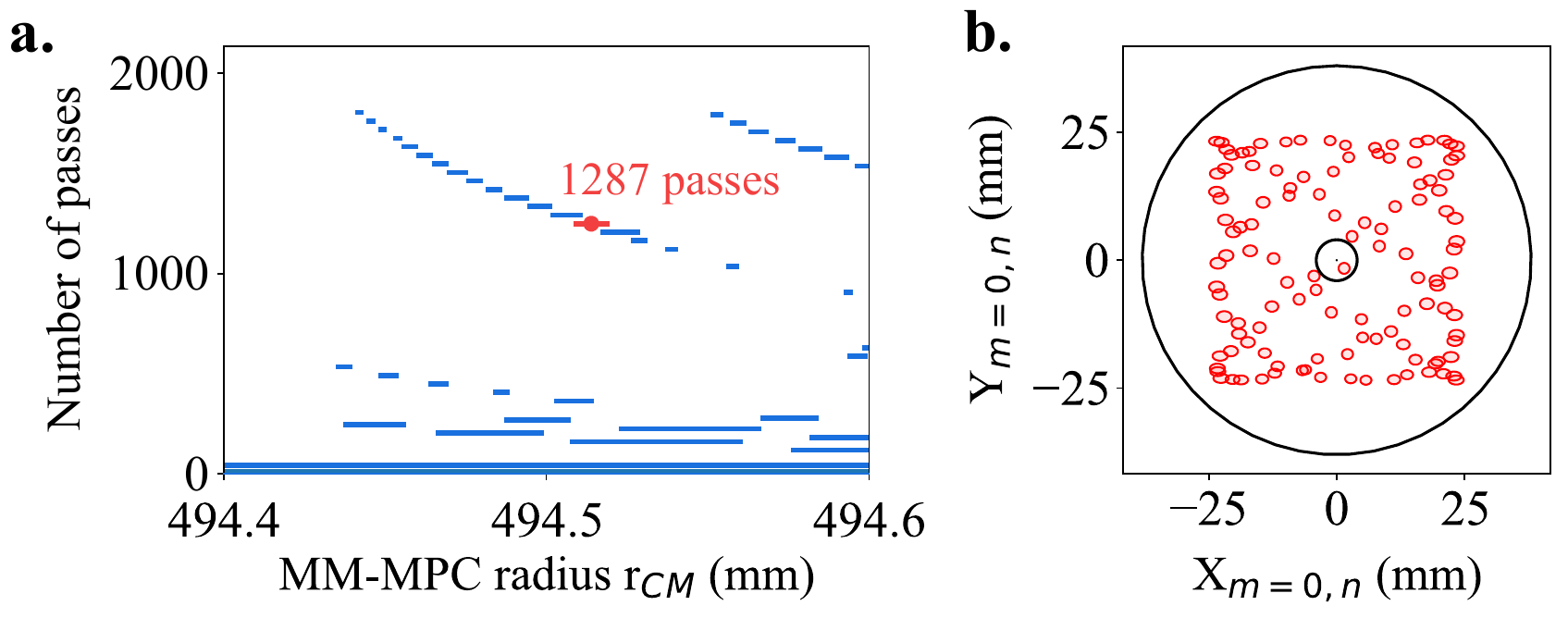}
		\caption{\textbf{a.} Calculated number of passes of non-clipping patterns for different radius of the 11-mirror MPC with 3-inch diameter mirrors. The hole diameter is $d_h=$6.6 mm, optimised for a pattern with 1287 passes. \textbf{b.} Corresponding simulated pattern, where the ellipses represent the beam size at 1/e$^2$.}
		\label{fig6:1287patterns}
	\end{figure}
	
	To reach higher pass numbers with the same 11-mirror configuration, we numerically investigated the use of 3-inch diameter mirrors. This setup supports non-clipping patterns with more than 1300 reflections (Figure~\ref{fig6:1287patterns}a).
	
	The pattern presented in Figure~\ref{fig6:1287patterns}b is calculated for $r_{CM}$=495.491 mm and supports 1287 passes. The tolerance of this pattern on $r_{CM}$ is 10 $\mu$m, assuming a perfect alignment of the mirrors. In practice, slight adjustments to the angle of mirror $m=M-1$ allow for fine-tuning of the beam position on mirror $m=0$, effectively increasing the experimental tolerance on $r_{CM}$. Additionally, off-centering the coupling hole or the input beam can further optimize this tolerance, making the pattern experimentally achievable.\par
	
	\subsection{Spectral broadening in air}
	
	\begin{figure}[t]
		\centering
		\includegraphics[width=\linewidth]{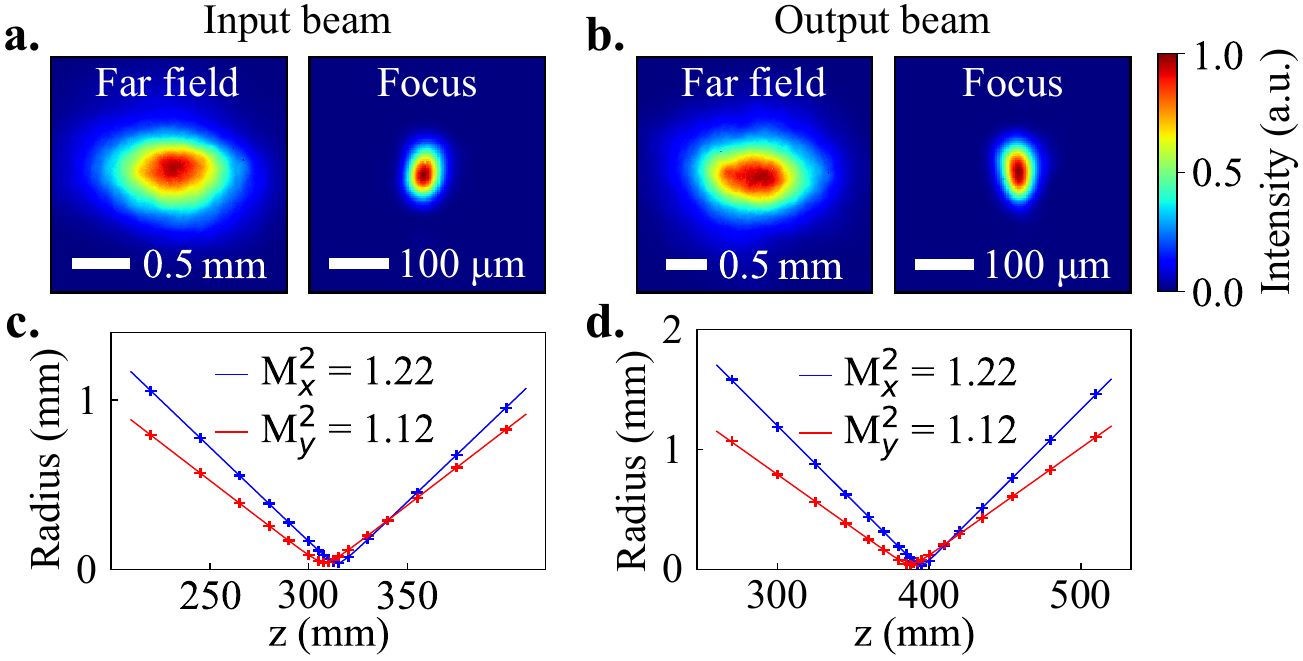}
		\caption{Far field and focused beam profiles of \textbf{a.} input and \textbf{b.} output beams of the burst-mode laser at low input pulse energy. M$^2$ factor measurement of \textbf{c.} input and \textbf{d.} output beam at low input pulse energy.}
		\label{fig5:profiles}
	\end{figure}
	
	To further investigate the spectral broadening concept, we used a burst-mode Yb:YAG laser \cite{seidel_factor_2022}, which was reconfigured to emit about 1\,ps pulses with up to 1\,mJ pulse energy at 1030\,nm, and a Pharos laser (Light Conversion) emitting 210\,fs pulses with up to 200\,$\mu$J pulse energy. A high-energy long-pulse laser was not available at the time of the experiments. The transmission of the 297-pass MPC decribed in section~\ref{sec:11MMMPC} was about 88\,\%, in good agreement with the specified reflectance of the used dielectric mirrors (Optoman, $R > 99.95\,\%$). First, we investigated if the introduced astigmatism deteriorates the beam quality of the laser. With the burst-mode laser operated at low power, we measured for both the input beam (Figure~\ref{fig5:profiles}a) and the output beam (Figure~\ref{fig5:profiles}b) an M$^2$-factor of 1.2 in x- and 1.1 in y-direction and nearly the same astigmatism. There is no additional astigmatism at the output, and the angle of incidence on the mirrors is small enough to avoid beam degradation from other aberrations such as coma.
	Moreover, we studied spectral broadening in air. In recent MPC experiments with air as nonlinear medium, spectral broadening factors of about five for 670\,fs, 3\,GW peak power pulses \cite{omar_spectral_2023} and of about 20 for 1.1\,ps, 7\,GW peak power pulses \cite{schonberg_compact_2025} have been reported. We obtained a broadening factor of 15 with pulses of only about 260\,MW peak power (Figure~\ref{fig6:spectra}a), indicating the great potential of MM-MPCs to achieve ultrahigh B-integrals. Similar to refs.~\cite{omar_spectral_2023,schonberg_compact_2025}, we observed a pronounced peak around 1030 nm, and a red-shifted asymmetric spectrum which is due to the Raman response of air. When using the 210\,fs input pulses, the 1030\,nm peak was not visible in the broadened spectra, probably due to better higher-order chirp management of the laser. Already at 25\,$\mu$J input energy ($\approx$110\,MW peak power), a minimum FT limit of about 30\,fs was reached (Figure~\ref{fig6:spectra}b). At higher input pulse energies, spectral components extended beyond the reflection band of the mirrors, leading to losses and an increase in the $M^2$-factor.\par 
	
	\begin{figure}[t]
		\centering
		\includegraphics[width=\linewidth]{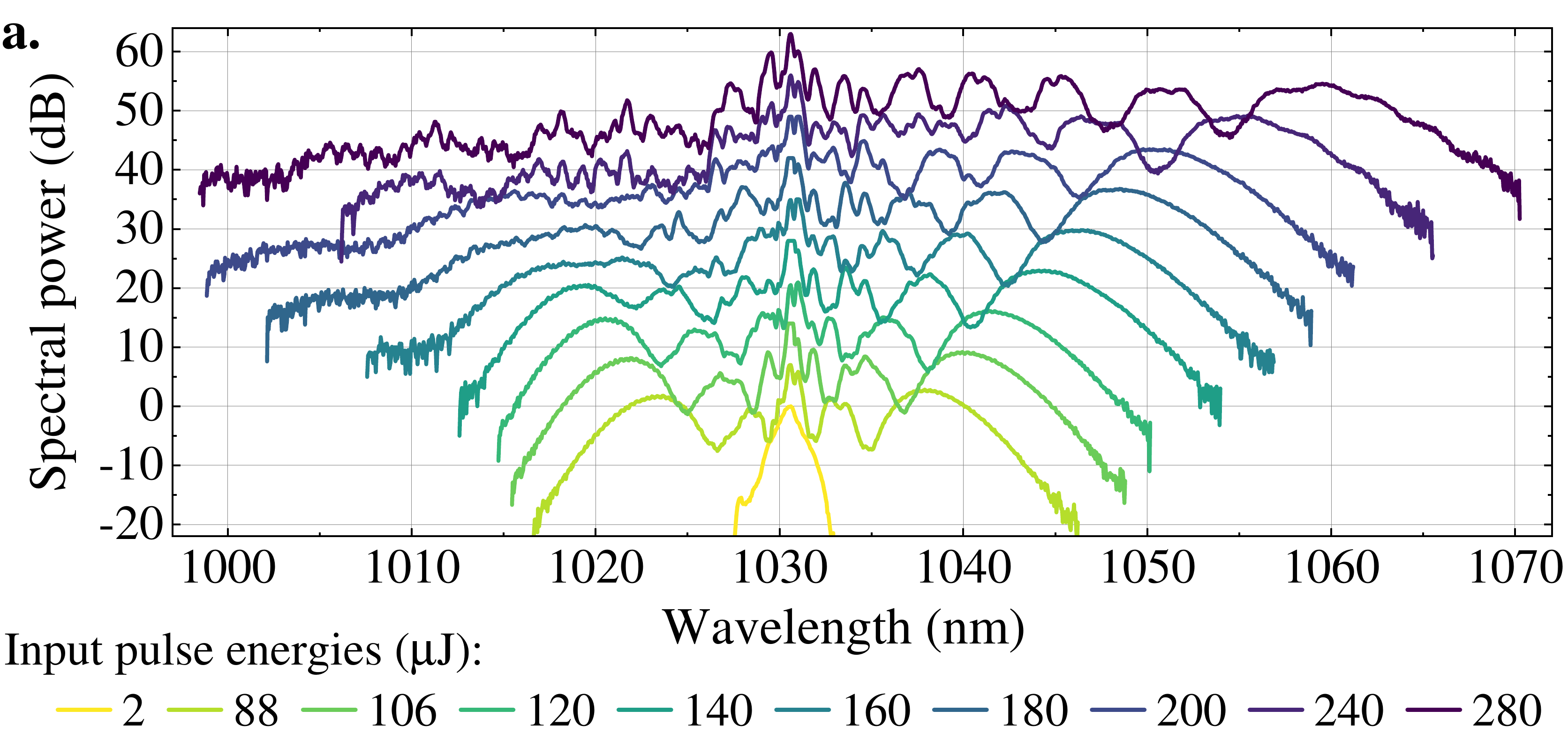}\\
		\includegraphics[width=\linewidth]{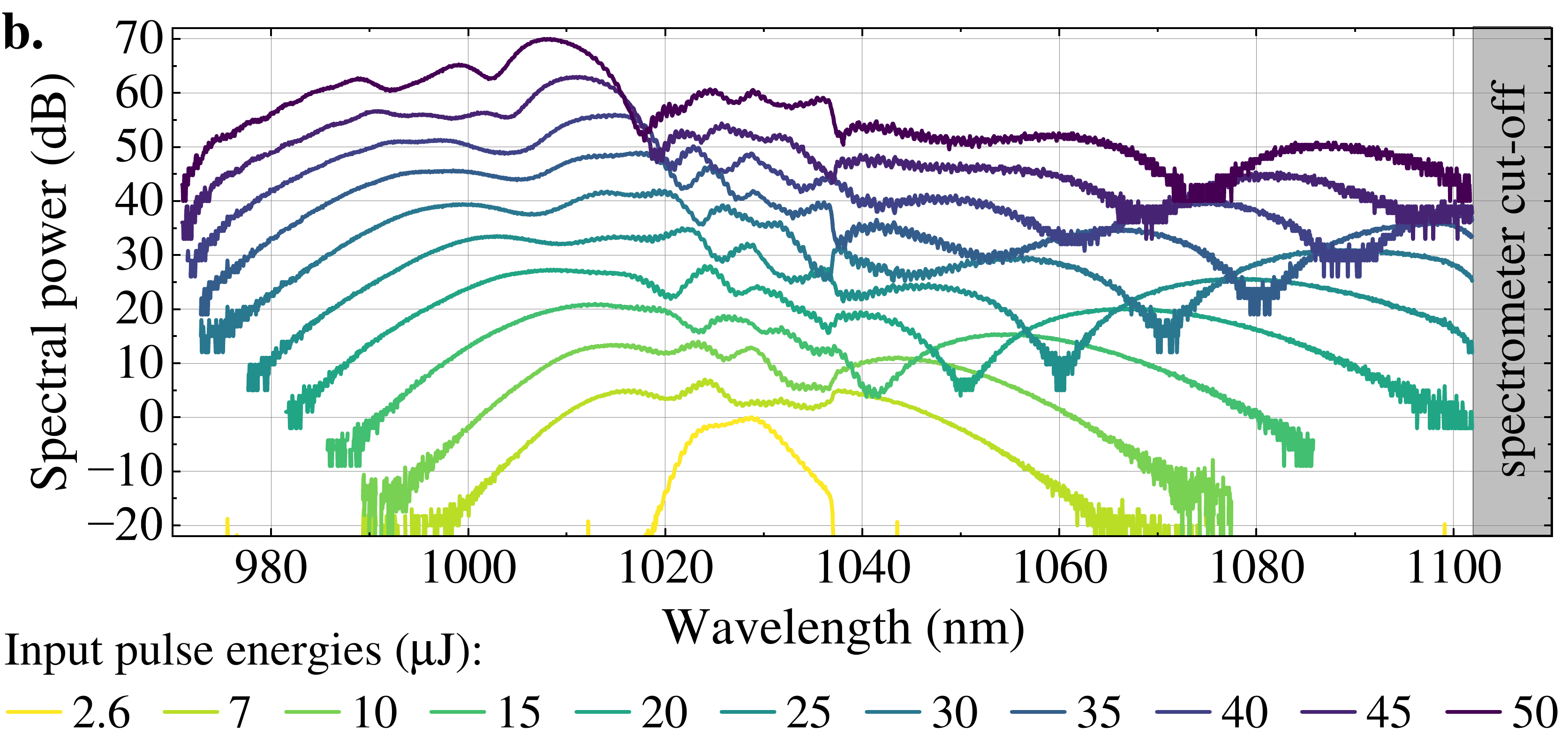}
		\caption{Spectral broadening with a 297-pass MPC exploiting the nonlinearity of air. Experiments were performed with \textbf{a.} a 1\,ps burst-mode Yb:YAG laser and \textbf{b.} a 210\,fs Pharos laser. The input energies are displayed in the legends. The spectra are offset by 7\,dB for clarity.}
		\label{fig6:spectra}
	\end{figure}
	
	\begin{figure*}[t]
		\centering
		\includegraphics[width=0.9\textwidth]{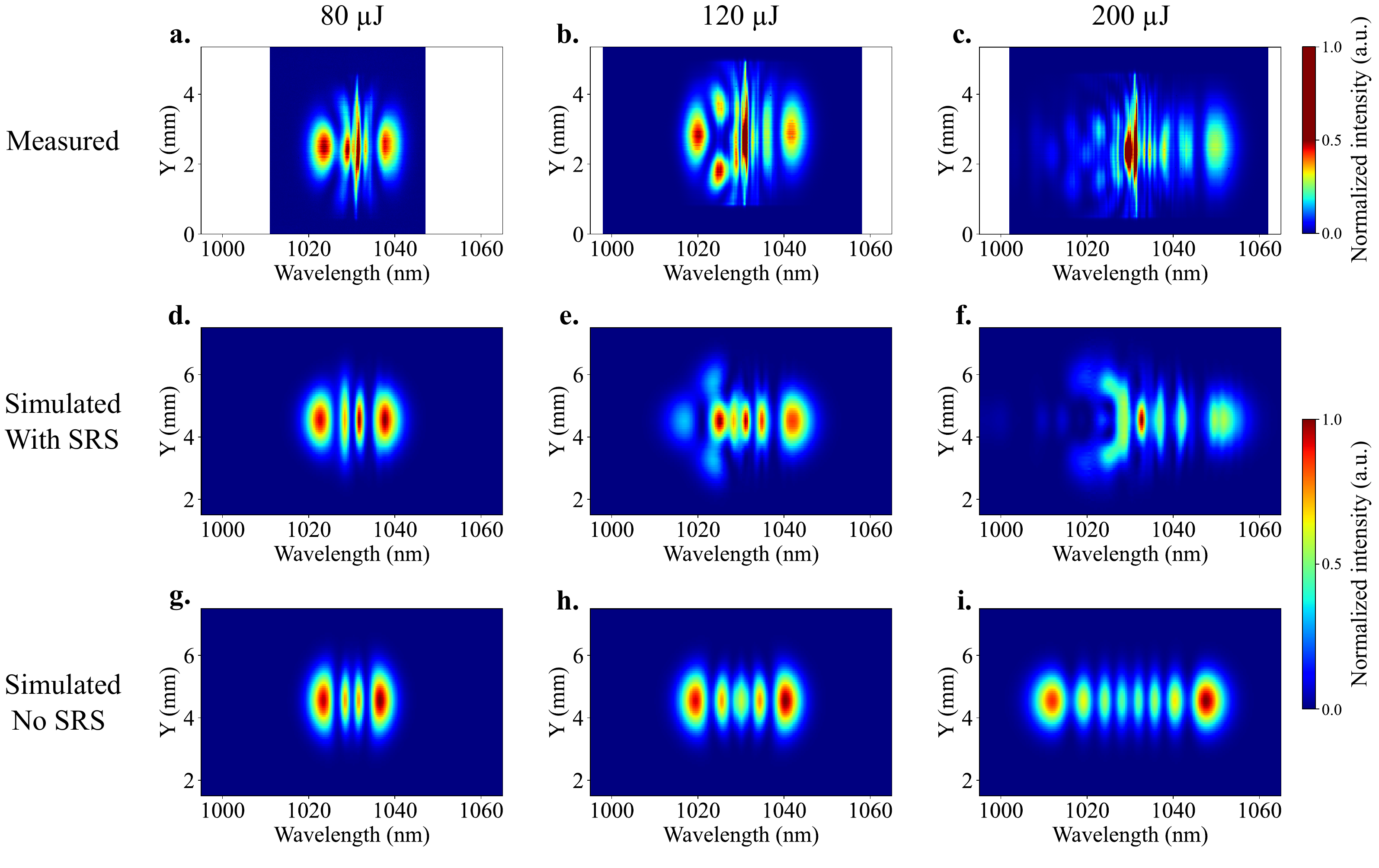}
		\caption{Measured (\textbf{a.}-\textbf{c.}) and simulated (\textbf{d.}-\textbf{i.}) spatio-spectral couplings at the output of the MM-MPC after 297 passes, using the burst-mode laser at different input pulse energies. The colormap of the measurements is normalized such that the strong peak at 1030 nm is saturated in order to improve the visibility outside the main peak. The left column is for 80 \textmu J input pulse energy, the central one for 120 \textmu J and the right one for 200 \textmu J. The simulations on the second line (\textbf{d.}-\textbf{f.}) are performed for a Raman fraction of $f_R$=0.77. The simulations on the third line (\textbf{g.}-\textbf{i.}) are performed without SRS by setting $f_R$=0.}
		\label{fig8:Raman_Simulations}
	\end{figure*}
	
	\subsection{Beam degradation from stimulated Raman scattering in air}
	
	A significantly stronger dependence of the $M^2$-factor on input pulse energy was observed in the spectral broadening experiments using the burst-mode laser. At an input energy of 200\,\textmu J, M$^2$ values of $3.2\times3.1$ were measured. Such high M$^2$ values do not allow for a higher-order Gaussian interpretation of the beam quality. To further investigate this degradation, we analyzed spatio-spectral couplings in the output beam using an ANDOR Kymera 193 Czerny-Turner-type spectrograph. The collimated beam was sent through a 10 \textmu m horizontal slit and dispersed by a 500 lines/mm reflection grating located in the 2f plane. These measurements were conducted at different input pulse energies, and the results are plotted in Figure~\ref{fig8:Raman_Simulations}. The wavelength axis was calibrated by comparing the spatially integrated spectrum to a reference measurement from a grating spectrometer. The measurements reveal spatio-spectral couplings, particularly at wavelengths below 1030 nm, that increase with the input pulse energy.

	We attribute these couplings to stimulated Raman scattering (SRS) in air \cite{Watson:25}. When the spectrum is broad enough, SRS leads to the scattering of high-energy photons to lower energy photons, resulting in the red-shifted spectrum that can be observed in Figure~\ref{fig6:spectra}. With the additional astigmatism introduced by the non-zero angle of incidence on the spherical mirrors, new spatial frequencies are generated at the focus at higher wavelengths, causing the beam degradation. To confirm this, we performed numerical simulations using the SISYFOS package \cite{arisholm_simulation_2021}. SRS is accounted for in the definition of the $\chi^{(3)}$ tensor by including a delayed function corresponding to the rotational response of the molecules. The time-dependent nonlinear index is expressed as:
	
	\begin{equation}
		\Delta n(t) = n_2 (1-f_R)  I(t)  + n_2 f_R \int_{-\infty}^t \text{d} \tau I(\tau) R(t-\tau)
		\label{eq:n_raman}
	\end{equation}
	
	where $n_2$ is the nonlinear optical index, $f_R$ is the Raman fraction, $I$ the intensity, $R(t) = \frac{\Gamma^2+\omega_R^2}{\omega_R} \exp (-\Gamma t)\sin(\omega_R t) $ is the response function of the molecular gas, with $\Gamma$ the damping time of the induced molecular rotations and $\omega_R$ their frequency. The first term in Eq.~\eqref{eq:n_raman} represents the instantaneous electronic response and the second term accounts for the delayed Raman contribution. 
	The values taken from ref.~\cite{beetar_multioctave_2020,Reichert:16} for the simulation of the propagation in 1 bar of air are summarized in Table \ref{tab:simu_parameters}. Beam propagation was computed in the full {x,y} plane with $128\times128$ spatial grid points, and $2^9$ points in the time axis, over 297 passes. Astigmatism was accounted for by considering astigmatic mirrors of focal lengths $f_{tan}$ and $f_{sag}$. 
	
	\begin{table}[b!]
		\caption{Values taken from ref.~\cite{beetar_multioctave_2020,Reichert:16} used to simulate propagation in 1\,bar of air in the MM-MPC with the SISYFOS package.}
		\label{tab:simu_parameters}
		\centering
		\begin{tabular}{cccc}
			\hline
			$n_2$  & $f_R$ & $\Gamma^{-1}$ & $\omega_{R}$ \\ 
			\hline
			$3.2\times10^{-19}$cm$^2$/W & 0.77 & 20 ns & 10 THz \\
			\hline
			
		\end{tabular}
	\end{table}
	
	The simulations with SRS presented in Figure~\ref{fig8:Raman_Simulations} reproduce the experimentally observed couplings in the short-wavelength region of the spectrum. While some discrepancies are present, likely due to the idealized input Gaussian pulse and beam, and to the simplified single-oscillator Raman model of air, the simulations qualitatively agree with the measurements.  
	When turning off SRS by setting the Raman fraction $f_R$ to 0, the spatio-spectral couplings disappear for every input pulse energy. This confirms that SRS is the dominant mechanism behind the observed beam degradation. Consequently, using a noble gas as the nonlinear medium can effectively eliminate these couplings, since SRS is not present, and preserve beam quality even at high pulse energies. 
	
	\section{Conclusions}
	In summary, we verified the CAA concept by rigorous simulations which showed that massive spectral broadening can be achieved in an MPC supporting 650 roundtrips. We found that sub-critical self-focusing and under certain conditions astigmatism are limiting the attainable compression factors. Moreover, we have demonstrated that multi-mirror geometries can provide km-scale nonlinear propagation lengths which are essential for boosting single-stage spectral broadening factors. In a proof-of-concept experiment, we demonstrated 297 passes through an eleven-mirror MPC and showed the excellent potential of extreme SPM by spectral broadening experiments in air. Nevertheless, multiple experimental milestones lie ahead on the road towards fully implementing the CAA scheme and will be subject of follow-up research: First, the use of an energetic multi-100-ps laser, second, the increase of passes inside the MM-MPC by enlarging the cell mirror diameter from two to three inches, third, the use of an atomic gas as nonlinear medium and finally, pulse shaping to improve the temporal contrast of the strongly self-phase modulated pulses after compression. \par
	We restricted our numerical investigations to the test case of a 1-m MPC diameter and pulses with 100\,mJ energy as well as 300\,ps FWHM duration. Similar to two-mirror MPCs, energy scaling can be accomplished by increasing the MPC mirror distance although the optical damage mechanism for sub-ns pulses may restrict scalability by beam area enlargement. Alternatively, attainable mirror reflectances of 99.999\,\% would allow to propagate the pulses even over 10\,000 passes. Longer propagation distances would allow longer input pulses with the same peak power but more energy. Finally, if efficient peak power scaling concepts for ultrashort pulse MPCs are experimentally verified at sub-TW peak power levels, double-stage compression from a few ns duration to the 100-fs range may become a viable generation scheme for Joule-class ultrafast lasers.\par
	The CAA scheme we proposed addresses four-level laser media with good energy and power handling properties but limited intrinsic emission linewidths. Its full implementation may enable cost-efficient and multi-kHz repetition rate energetic femtosecond sources. Industrially highly mature laser technology may be efficiently converted to the ultrafast regime.
	
	\appendix{}
	
	\sloppy{}
	\section{Appendix}

\subsection{Comparison to spectral broadening of sub-ns pulses in hollow-core fibers}
\label{app1:hc-fiber}

We note that anti-resonant hollow-core fibers (AR-HCFs) could be a potential alternative to MPCs for the CAA approach. AR-HCFs can be coiled, and thus lead to extremely compact setups for TW-level pulse generation. They have also shown good average power handling capabilities, for instance, $> 80\,\%$ transmission over a 1\,km propagation length was reported at kW average power levels \cite{mulvad_kilowatt-average-power_2022}. However, only transmission of pulse energies up to a few mJs in single-mode and about 20\,mJ in multi-mode operation has been demonstrated to date \cite{lekosiotis_-target_2023,zhao_delivery_2024} although Joule-class pulse energy handling was predicted \cite{debord_hollow-core_2019}. The main advantage of AR-HCFs is the preservation of small beam sizes over the full nonlinear propagation distance which effectively leads to shorter nonlinear lengths in comparison to MPCs, and thus to larger spectral broadening factors for equal propagation distances. On the other hand, this leads also to the experimental challenge of mode-matching at intensities that exceed the damage threshold of the fiber glass by orders of magnitude which results in significantly lower mode-mismatch tolerances in comparison to MPCs. To apply the CAA approach to AR-HCFs,  dispersion, nonlinearity and input peak power must be carefully balanced in order to avoid ionization, losses that could damage the fiber glass and modulational instabilities. We show simulations of spectral broadening in AR-HCFs in Figure~\ref{fig:AR-HCF} for the rare gases krypton (a), argon (b) and helium (c). A mode-field diameter (MFD) of 44\,$\mu$m was chosen and transmission losses of 2\,dB/km assumed. We estimated the fiber dispersion by the hollow-core capillary equation \cite{travers_ultrafast_2011} and approximated the bore radius in the dispersion calculations by MFD/2. Moreover, only the beam averaged intensity was considered and polarization effects as well as fiber-design specific properties (e.g. fractional power in the glass cladding) were neglected, such that the simulation results should be interpreted as an upper limit for the prospects of CAA in AR-HCF.\par

\begin{figure}[t!]
	\centering
	\includegraphics[width=\linewidth]{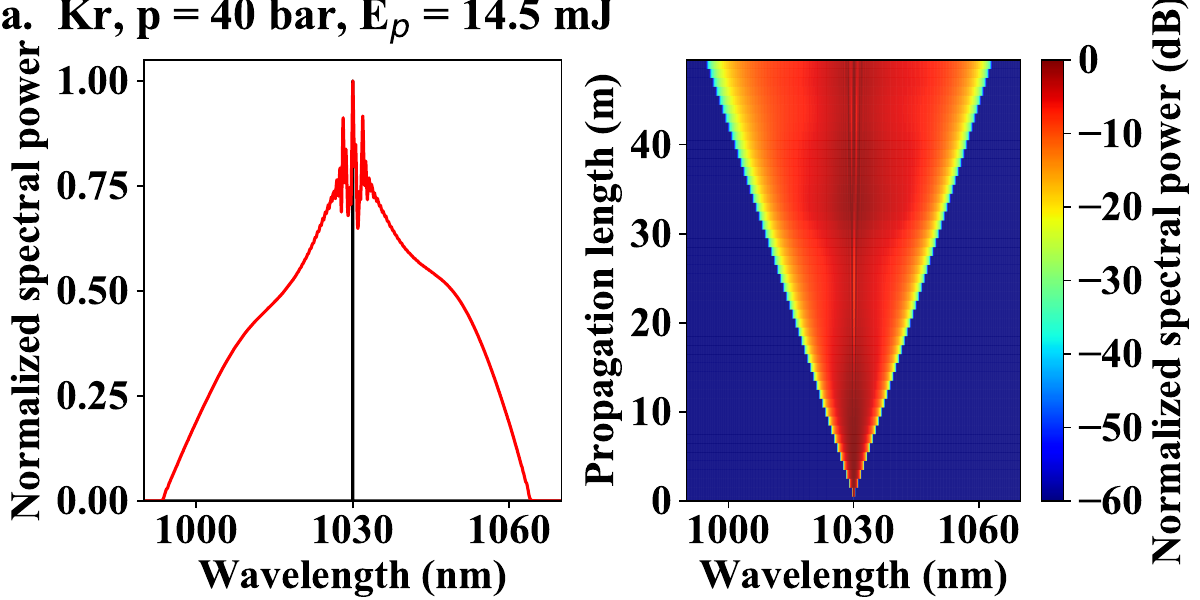}\\
	\includegraphics[width=\linewidth]{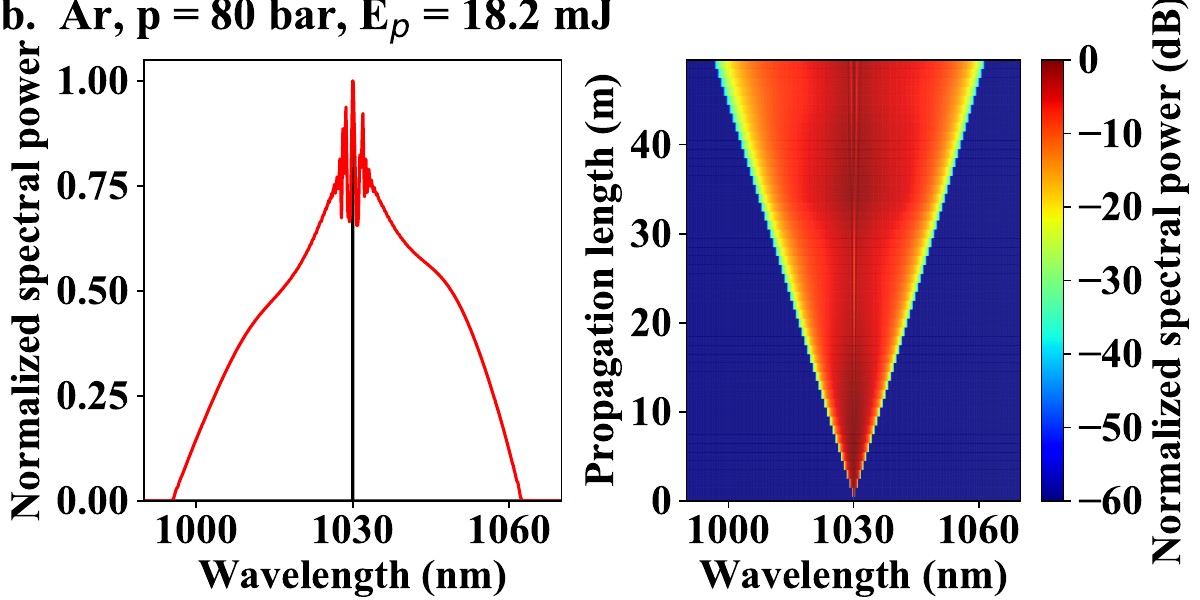}\\
	\includegraphics[width=\linewidth]{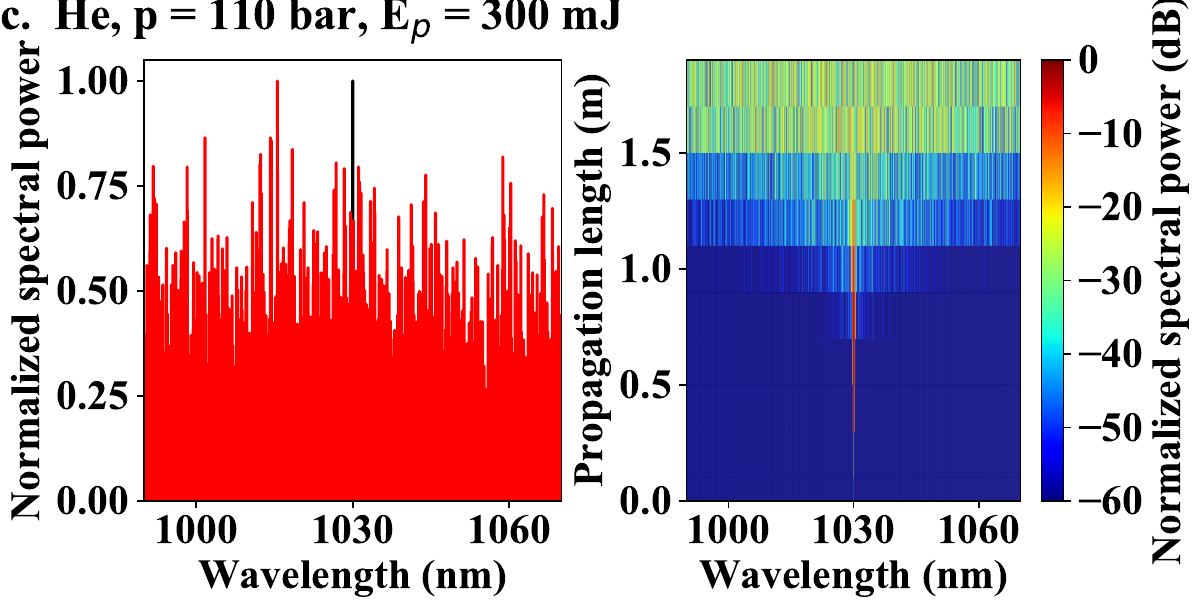}
	\caption{Numerical study of spectral broadening in hollow-core fiber with 44\,$\mu$J mode-field diameter for parabolic input pulses of 300\,ps FWHM duration. (\textbf{a.}) The krypton pressure of 40\,bar resulted in normal dispersion of 480\,fs$^2/$m at 1030\,nm and spectral broadening to a 51\,fs FT-limited duration after 50\,m of propagation. (\textbf{b.}) The argon pressure of 80\,bar resulted in normal dispersion of 460\,fs$^2/$m at 1030\,nm and spectral broadening to a 54\,fs FT-limited duration after 50\,m of propagation. (\textbf{c.}) The helium pressure of 110\,bar resulted in anomalous dispersion of -510\,fs$^2/$m at 1030\,nm and the inset of modulational instabilities after less than 1\,m of propagation}
	\label{fig:AR-HCF}
\end{figure}

As for the MM-MPC simulations, we considered parabolic pulses of 300\,ps FWHM duration. For Kr and Ar, the input pulse energy was limited by the onset of multi-photon ionization. We set the peak intensities to 6\,TW/cm$^2$ (Kr) and 7.5\,TW/cm$^2$ (Ar), corresponding to half of the MPI detection limit in ref.~\cite{lhuillier_multiply_1983}. Moreover, we set the pressure such that  $P_p\approx 0.2 P_{cr}$ in proximity to the optimal value derived in section~\ref{subsec:sim_compress}. The corresponding pressures of 40\,bar and 80\,bar, respectively, resulted in spectral broadening in the normal dispersion regime. For helium, to our knowledge, the highest experimentally used pressure of 110\,bar was set \cite{azhar_raman-free_2013}, which nevertheless resulted in anomalous dispersion in the fiber. The pulse energy was adjusted again to operate at $P_p\approx 0.2 P_{cr}$. The corresponding peak intensity was about 300\,TW/cm$^2$ being about a factor four below the observed onset of MPI in ref.~\cite{lhuillier_multiply_1983}. For Ar and Kr, strong spectral broadening was observed and the FT-limit could be reduced to sub-60\,fs after 50\,m of propagation. As opposed to this, modulational instabilities (MIs) generated a broadband, but incoherent spectrum if helium was used as nonlinear medium. This agrees with the expectation of nonlinearly propagating a long pulse in the anomalous dispersion regime \cite{russell_hollow-core_2014}. To access the normal dispersion regime with He, the mode-field diameter would have to be strongly enlarged. However, this would in turn lead to a massive increase of microbending losses \cite{fokoua_loss_2023}. Therefore, starting with a few-ns pulses while keeping Ar or Kr as nonlinear medium appears to be the better approach to increase the pulse energies beyond the 20\,mJ level, under the premise that the experimental results further approach the theoretical predictions of deliverable pulse energy.\par
We note that MIs emerged through semi-classical noise in the simulations. This was numerically implemented by adding half a photon to each spatio-temporal mode. Signatures of MIs were absent in all of the conducted MPC simulations owing to the generic operation in the normal dispersion regime. \par

\begin{figure*}[t!]
	\centering
	\includegraphics[width=.95\textwidth]{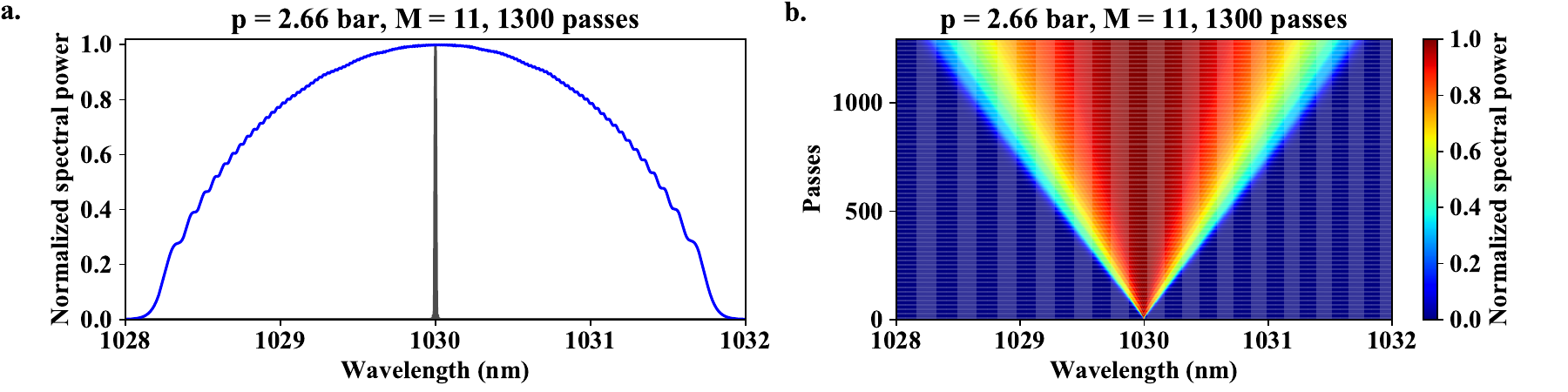}
	\includegraphics[width=.95\textwidth]{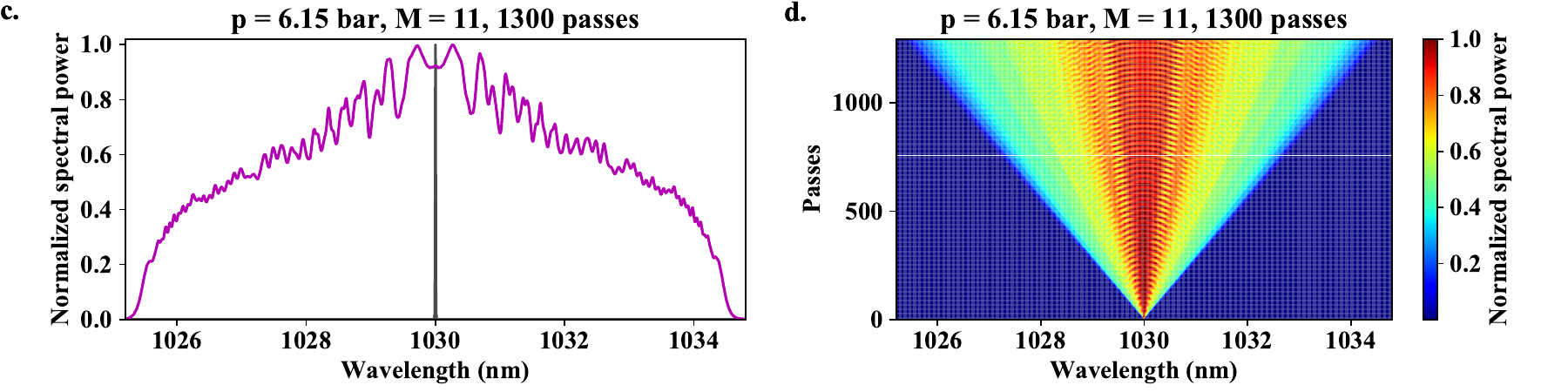}
	\includegraphics[width=.95\textwidth]{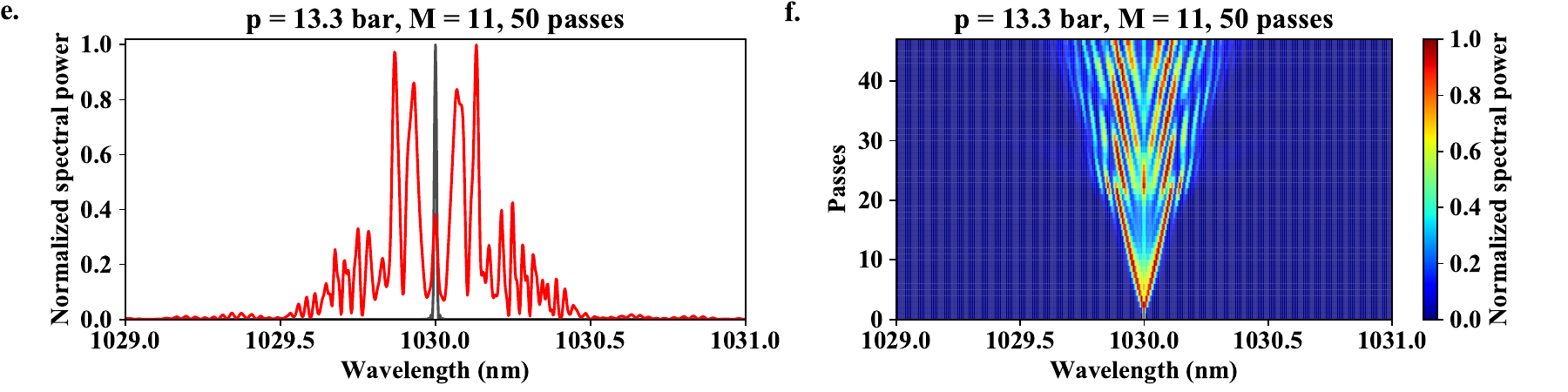}
	\includegraphics[width=.95\textwidth]{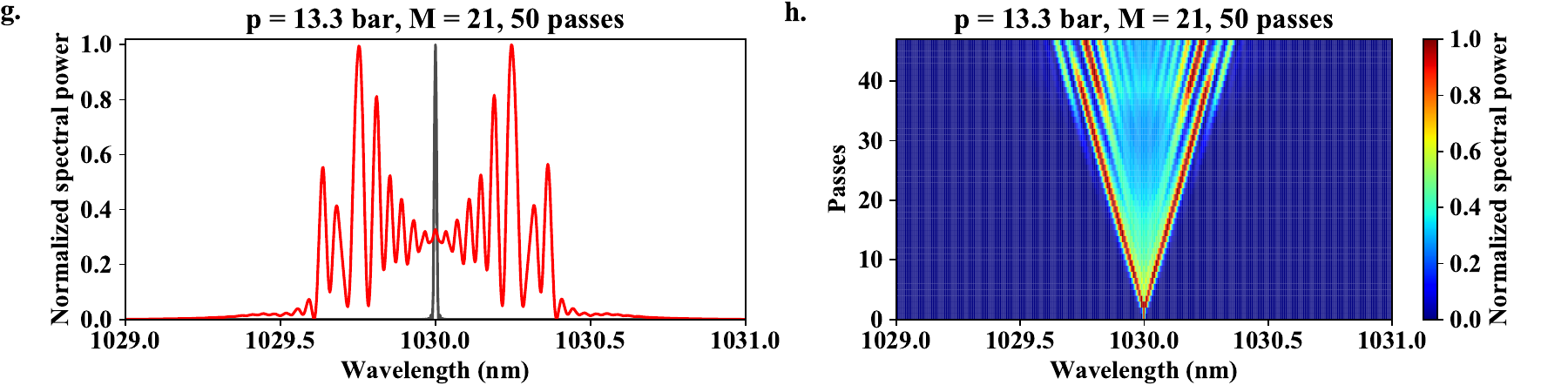}
	\caption{Simulations of spectral broadening in MM-MPCs including the astigmatism originating from the oblique angles of incidence on the MPC mirrors which were calculated by 90$^\circ$/M. \textbf{a., c., e., g.} show the spectra before (black lines) and after spectral broadening in an 11-mirror MPC (blue, violet and red lines) for different krypton pressures (a.,c.,e.) and in a 13-mirror MPC for $p = 6.15$\,bar  (g.). The false color plots \textbf{b., d., f., h.} show the evolution of the spectral broadening over the simulated passes. For $p = 2.66\,$bar (b.), it resembles plane wave propagation. Only minor modulations of the spectrum are visible. For $p = 6.15\,$bar (c., d.) periodic modulations of spectrum appear. These are getting very strong for $p = 13.3\,$bar. Using a larger number of MPC mirrors and smaller angles of incidence, respectively, leads to a moderate reduction of the modulations (g., h.). They are, however, still much stronger than for $p = 6.15\,$bar.}
	\label{fig:astig_broadening}
\end{figure*}

\subsection{Simulation of astigmatic multi-pass cells}
\label{app2:astig}

For investigating the impact of moderate astigmatism which the experimentally studied MM-MPCs induce, simulations with a 129 x 129 $x$-$y$-grid and a reduced time grid with up to $2^{13}$ points were conducted. Due to the mirror symmetry with respect to the $x=0$ and $y=0$ axes, only the quadrant with positive $x-$ and $y-$values was simulated. The focal lengths of the curved MPC mirrors were set to $f_{sag}= \frac{\mathcal{R}}{2}\cos(\pi/(2M))$ in the sagittal plane and $f_{tan}= \frac{\mathcal{R}}{2\cos(\pi/(2M))}$ in the tangential plane, where $\mathcal{R}$ is the radius of curvature of the mirrors and $M$ the number of mirrors in the MM-MPC. All other parameters were kept as before. Only parabolic input pulses described by Eq.~\eqref{eq:parabol} were investigated. We looked at different input beam settings: On the one hand, round beams and mode-matching under the assumption $\cos(\pi/(2M)) \approx 1$ were used. On the other hand, elliptical beams with radii of curvature at the first MPC mirror matched to $\mathcal{R}\cos(\pi/(2M))$ and $\frac{\mathcal{R}}{\cos(\pi/(2M))}$, respectively, were used. While the latter approach reduced the intensity fluctuations on the mirrors, differences in the self-phase-modulated spectra were not observed. Particular attention was paid to the interplay between astigmatism and the critical-to-peak-power ratio. For an $M=11$-mirror MPC and $P_{cr}/P_p=10$, spectral broadening resembled the propagation without transversal degrees of freedom as shown in Figure~\ref{fig:astig_broadening}a,b. The pulses passed the astigmatic MPC 1300 times which resulted in FT limited pulse duration of 965\,fs. The simulated spectra for $P_{cr}/P_p=4$ (Figure~\ref{fig:astig_broadening}c,d) are qualitatively different from the flat-top spectrum obtained at $p = 6.15$\,bar under the assumption of cylindrical symmetry (Figure~\ref{CAA_sim_press}a, violet line). After including astigmatism, the spectrum exhibits clear modulations near the central wavelength. For $p = 13.3$\,bar, i.e. $P_{cr}/P_p=2$, the spectral modulations became very strong already after about 20 passes (Figure~\ref{fig:astig_broadening}e,f). Figs.~\ref{fig:astig_broadening}g,h show that, to some extent, the spectral modulations are reduced by increasing the numbers of mirrors in the MPC and by reducing astigmatism, respectively. However, the modulations are still much stronger than for the $P_{cr}/P_p=4$ simulation despite the smaller spectral broadening. In summary, the simulations predicted that small $P_{cr}/P_p$ ratios, on the one hand, impede the compression of the broadened spectrum by dispersion compensation, and on the other hand, lead to strongly modulated spectra which do not support high pulse contrast. Therefore, $p = 13.3$\,bar was not considered for pulse compression simulations.\par

\begin{figure*}[t!]
	\centering
	\includegraphics[width=\textwidth]{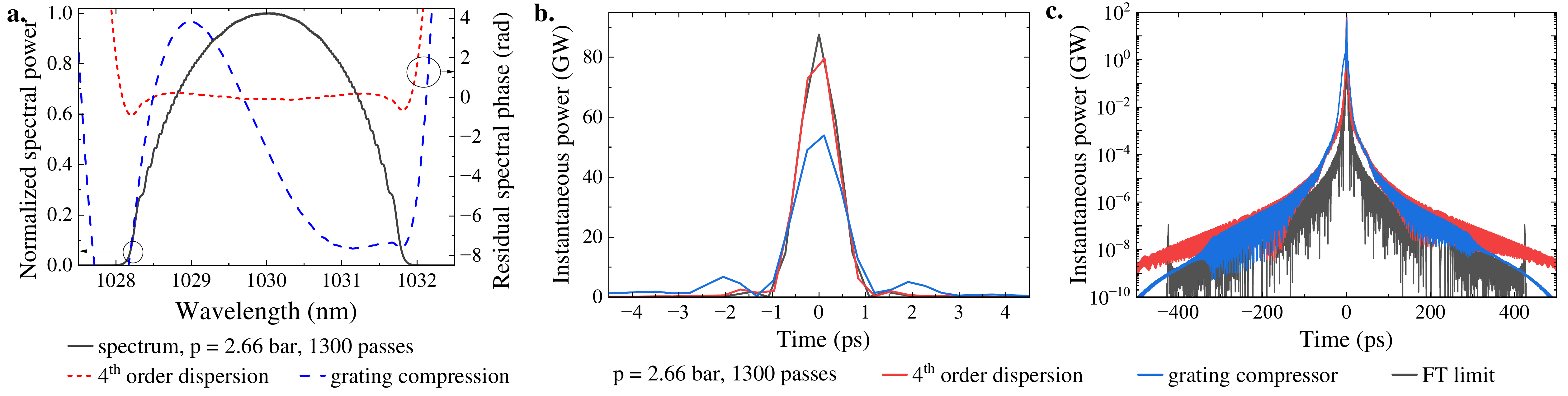}
	\includegraphics[width=\textwidth]{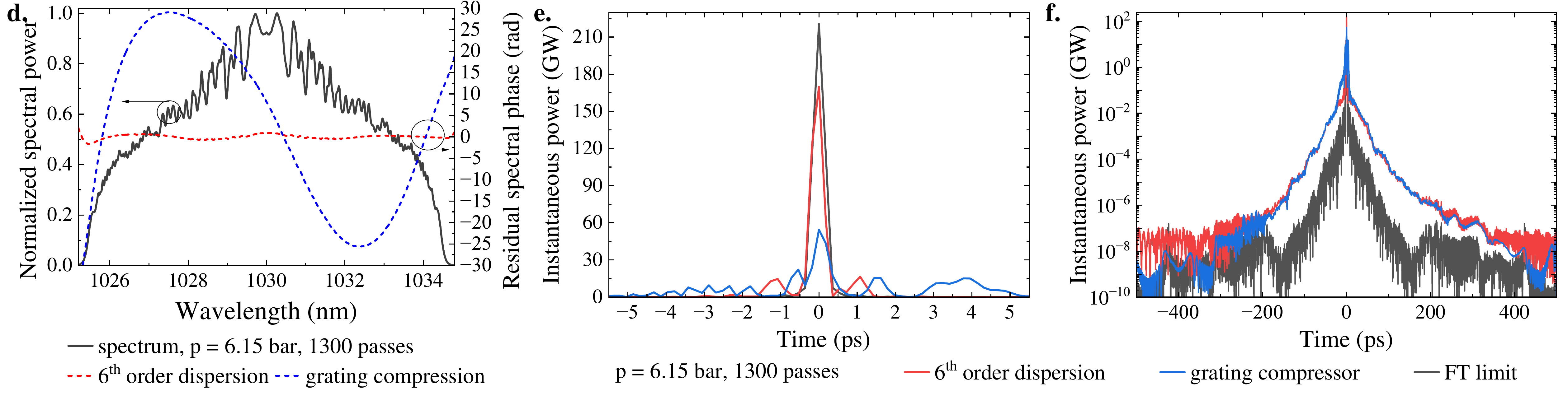}
	\caption{Simulated compression of spectrally broadened pulses with parabolic envelop at $p = 2.66\,$bar (\textbf{a.-c.}) and $p = 6.15\,$bar (\textbf{d.-f.}) The blue lines show the results for applying a grating pair compressor, the red lines for applying a phase shaper with full control over higher-order dispersion. The optimized grating pair distances were 1.790\,m for $p = 2.66\,$bar and 0.702\,m for $p = 6.15\,$bar. The controlled dispersion orders of the phase shaper were increased until near FT limited pulses could be achieved. For $p = 2.66\,$bar, the dispersion terms up to the fourth order were controlled, being -60.0\,ps$^2$, -28.6\,ps$^3$ and -1.91\,ps$^4$. For $p = 6.15\,$bar, the dispersion terms up to the sixth order were controlled, being -25.26\,ps$^2$, -0.11\,ps$^3$, 0.62\,ps$^4$, -4$\times10^{-3}\,$ps$^5$ and -0.19\,ps$^6$. In \textbf{a.} and \textbf{d.} the simulated spectra and residual spectral phases of the x = 0, y = 0 grid point after applying the compressor are shown. The corresponding pulses are plotted on a linear scale in \textbf{b.}, \textbf{e.} and on a logarithmic scale in \textbf{c.}, \textbf{f.}}
	\label{figSI:compress}
\end{figure*}

\subsection{Pulse compression for $p = 2.66\,$bar and $p = 6.15\,$bar pressures in the MM-MPC}
\label{app3:compress}

We show the compression of the pulses after propagation through the MM-MPC in more detail for the pressures $p = 2.66$\,bar and $p = 6.15$\,bar. We restrict the discussion to two compressor types: First, the GDD and TOD of a pair of gratings with 1500 lines/mm in Littrow configuration was used ($GDD =-34\,ps^2$,	$TOD=0.221\,ps^3$, for 1\,m grating separation). Second, a phase shaper with independent control of the dispersion orders was simulated. For $p = 2.66$\,bar, the application of a grating pair dispersion resulted in a compression that yielded more than 60\,\% of the FT limited peak power (Figure~\ref{figSI:compress}a-c, blue lines). Apart from two side lobes at $\pm 2$\,ps delay from the main peak, a very good pulse contrast was computed. Near FT-limited pulses and a flat phase over the broadened spectrum was obtained by additional control over the third and fourth order dispersion (red lines). Pulse compression became significantly more difficult for $p = p_{cr}/2 = 6.15$\,bar (Figure~\ref{figSI:compress}d-f). By using the second and third order dispersion of a grating pair, only about 25\,\% of the FT limited peak power could be achieved (Figure~\ref{figSI:compress}e). Furthermore, additional full control over third and fourth order dispersion did not significantly improve the compression quality. Only if the dispersion terms up to the sixth order were controlled, a nearly FT limited pulse could be computed. The pulse contrast over a 1\,ns time window was good for both grating compression and higher order phase control (Figure~\ref{figSI:compress}f).\par
	
	\subsection{Analytical solution to the propagation in the MM-MPC}
	
	\begin{figure*}[t!]
		\centering
		\includegraphics[width=\textwidth]{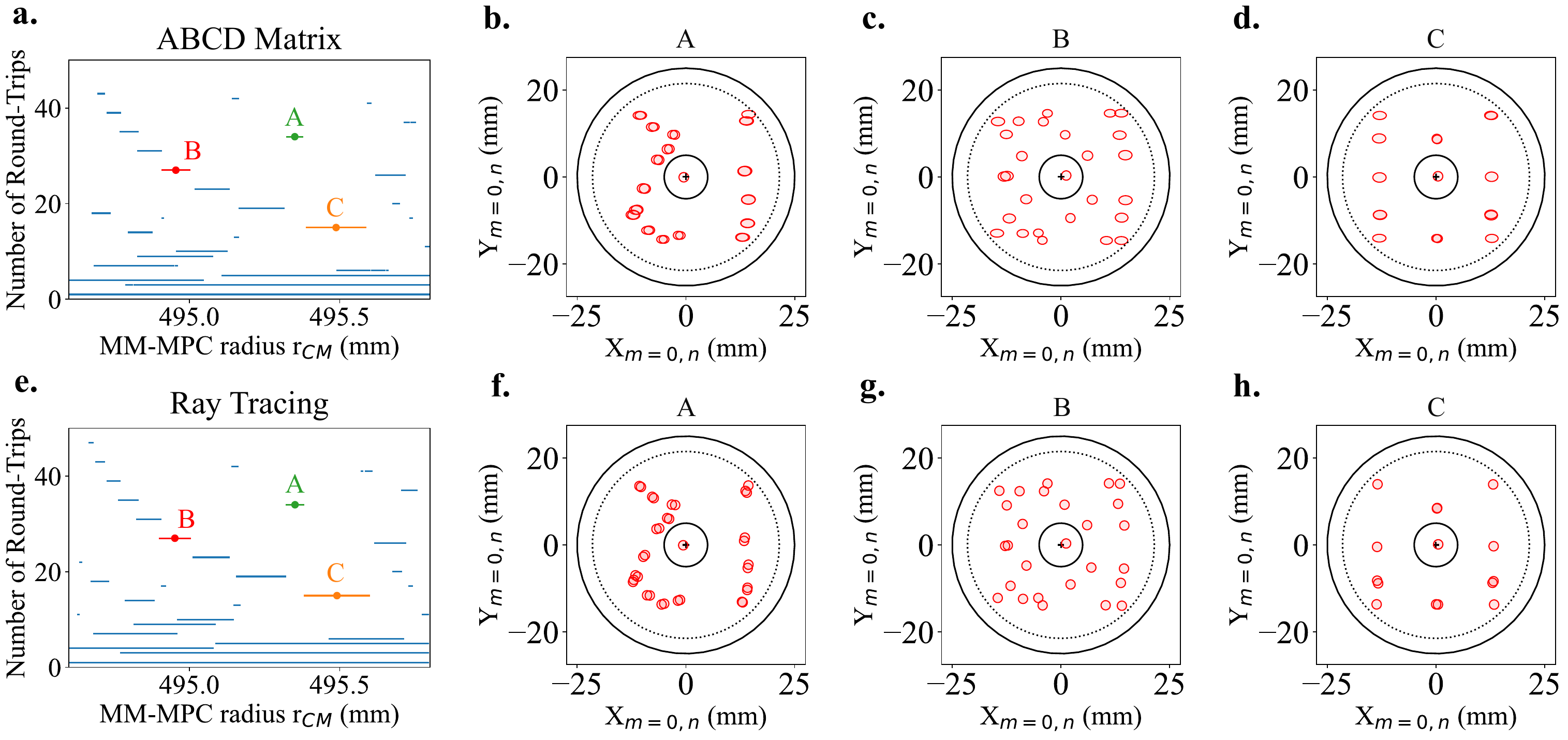}
		\caption{Calculated number of round-trips of non-clipping patterns
			for different radius of the 11-mirror MPC using \textbf{a.} the analytical solution from the ABCD matrix approach, and \textbf{e.} the ray tracing approach from ref.~\cite{kong_optical_2022}. Three selected patterns A, B and C with 374, 297 and 165 total passes respectively, (\textbf{b.},\textbf{c.},\textbf{d.}) obtained with the ABCD matrix approach, and corresponding patterns (\textbf{f.},\textbf{g.},\textbf{h.}) obtained with the ray tracing approach. The spots are projected onto the plane of the mirror m=0 for plotting, and the ellipses represent the beam size at 1/e$^2$. The dashed circles represent the mirror diameter $d_m$ that was taken in the clipping conditions, smaller than the actual size of the mirror to take into account the actual coated surface on the mirror specified by the manufacturer.}
		\label{figS3:comparison}
	\end{figure*}
	
	In order to calculate the beam trajectory through the MM-MPC, we employed the ABCD matrix formalism used for 2-mirror astigmatic MPCs \cite{herriott_folded_1965}. For an MPC composed of two slightly astigmatic mirrors with focal lengths $f_x$ and $f_y$ in two perpendicular directions, we can consider an equivalent resonator as a sequence of equally spaced thin lenses. It can then be shown that the x-y coordinates of the beam at the equivalent lens $i$ are given by:
	
	\begin{subequations}
		\begin{equation}
			X_i = X_{max} \sin(i\theta_x + \alpha),
			\label{eqS2:X}
		\end{equation}
		\begin{equation}
			Y_i = Y_{max} \sin(i\theta_y + \beta).
		\end{equation}
		\label{eqS2:XY}
	\end{subequations}

	where $X_{max}$, $Y_{max}$, $\alpha$ and $\beta$ are constants that depend on the injection condition, and the frequencies $\theta_x$ and $\theta_y$ are given by $\cos{\theta_x} = 1 - \frac{D}{2f_x}$ and $\cos{\theta_y} = 1 - \frac{D}{2f_y}$, $D$ being the distance between the center of the two mirrors.
	
	For the MM-MPC, as long as the angle of incidence on the mirrors remains sufficiently small, the off-axis spherical mirrors can be approximated as astigmatic mirrors with $f_{sag}\approx \frac{\mathcal{R}\cos(\varphi)}{2}$ in the sagittal plane and $f_{tan}\approx \frac{\mathcal{R}}{2\cos(\varphi)}$ in the tangential plane \cite{jenkins_fundamentals_1976}. Eq.~\eqref{eq:XYkm} is another writing of Eq.~\eqref{eqS2:XY}, where the position on each equivalent lens $i$ is attributed to a reflection $n$ on a mirror $m$. In Eq.~\eqref{eq:XYkm}, the constants $\alpha$ and $\beta$ are set to 0 since the beam is coupled in through a hole at the center of the mirror $m=0$.
	
	The beam radii on the mirror in the sagittal ($w_{m,sag}$) and tangential ($w_{m,tan}$) planes are calculated for each reflection. To do this, we use the transformation of the complex beam parameter by the ABCD matrices corresponding to free space propagation and reflection on the mirrors, with focal lengths $f_{sag}$ and $f_{tan}$, respectively. The initial condition is that the curvature of the input beam matches the radius of curvature of the mirror at the position of the hole, ensuring mode-matching for an MPC of 2 mirrors, in the linear case. The beam radius in the nonlinear mode-matching case will be smaller on the mirrors \cite{hanna_nonlinear_2021}, and the linear mode-matching value that we consider corresponds to an upper limit.
	
	To select only patterns that fully couple out of the MM-MPC, we impose certain conditions to avoid clipping. Clipping can occur either at the mirror edges or at the injection hole on the mirror m=0. A beam is considered clipped if the distance from its center to an edge is smaller than the beam radius on the mirror $w_m$. As the beam is elliptical because of the astigmatism of the mirrors, we take the biggest of the two radii for a given reflection: $w_m=\text{max}(w_{m,sag};w_{m,tan})$. This condition ensures that at least 86.5\,\% of the beam energy is reflected. The condition that the beams do not clip on the mirror of diameter $d_m$ reads:
	
	\begin{equation}
		X_{m,n}^2+Y_{m,n}^2<\left(\frac{d_m}{2} -w_m \right)^2.
	\end{equation}
	
	On the mirror $m=0$, there is the additional condition that the reflections must not clip on the hole of diameter $d_h$, and that the beam that couples out is fully transmitted through the hole. These conditions read, respectively:
	
	\begin{subequations}
		\begin{equation}
			X_{m=0,n}^2+Y_{m=0,n}^2>\left(\frac{d_h}{2} +w_m\right)^2,
		\end{equation}
		\begin{equation}
			X_{m=0,N}^2+Y_{m=0,N}^2<\left(\frac{d_h}{2} - w_m\right)^2.
		\end{equation}
	\end{subequations}
	
	For each radius of the MM-MPC $r_{CM}$, we calculate patterns for different values of $X_{max}$ and $Y_{max}$, only keeping the patterns that satisfy the clipping conditions. In Figure~\ref{fig3:MM-MPC_experiment} and Figure~\ref{figS3:comparison}a, $r_{CM}$ was varied from 494.3 mm to 495.8 mm with a step of 1.2\,\textmu m. For each value of $r_{CM}$, $X_{max}$ and $Y_{max}$ were both independently varied with 40 values between 0 and $d_m/2$. 
	
	For 2-inch mirrors, we considered a useable diameter of $d_m$ = 43 mm, which is smaller than the diameter of the mirror of 50.8 mm to account for the specified area of the coating supplier. This is represented by the dashed circles in Figure~\ref{figS3:comparison}. For 3-inch mirrors we considered a useable diameter of 69.2\,mm, smaller than the actual diameter of the mirror of 76.2\,mm.

	\subsection{Validity of the ABCD matrix approach}
	\label{app5:ABCD}
	
	To verify the validity of this analytical approach under our MPC parameters, we compared our results to the 3D ray propagation formalism used in ref.~\cite{kong_optical_2022}. In this method, rays are treated as vectors, and their reflections are calculated by considering the full 3D surface of the spherical mirrors. Unlike our analytical approach, this method does not approximate the off-axis spherical mirrors as astigmatic mirrors, providing the exact reflection positions. However, it does not give an analytical formula for the positions of the rays on the mirrors, does not allow to directly estimate the beam radius on the mirrors, and requires more computational power. 
	
	The same clipping conditions were taken for the 3D ray tracing approach, but the beam radius $w_m$ was kept constant for each reflection. Figure~\ref{figS3:comparison} presents different patterns obtained from both approaches. In our regime, which is close to the concentric regime ($D_{M} \approx 2R$) and with 11 mirrors that give an angle of incidence $\varphi = 8.2^\circ$ on the spherical mirrors, both approaches give similar results. Given this agreement, and for the purpose of this proof-of-principle experiment, we opted for the analytical approach, as it enables much faster computations. This efficiency allows us to quickly calculate a large number of patterns without the need for an optimization algorithm.
	
	\begin{center}\textbf{Acknowledgement}\end{center}
	
	\noindent  We thank DESY, a member of the Helmholtz Association HGF, for the provision of experimental facilities. G. B. and A.-L. V. acknowledge the financial support from the Swedish Research Council (grant No. 2022-03519) and the Knut and Alice Wallenberg Foundation. Moreover, we thank Yujiao Jiang for fruitful discussions about simulating spectral broadening in AR-HCFs and John Travers for discussing practical limitations of AR-HCFs.

	\bibliographystyle{unsrt}
	\bibliography{Beaufort_MM_MPC}
	
\end{document}